\documentclass[apj]{emulateapj}

\usepackage{subfigure} 
\usepackage{graphicx}
\usepackage[english]{babel}
\usepackage{graphicx}
\usepackage{float}
\usepackage{floatflt}
\usepackage{rotating}
\usepackage{epsfig}
\usepackage{longtable}
\usepackage{amsmath}
\usepackage{amsbsy}

\usepackage{mathptmx}

\def\gtorder{\mathrel{\raise.3ex\hbox{$>$}\mkern-14mu
             \lower0.6ex\hbox{$\sim$}}}
\def\ltorder{\mathrel{\raise.3ex\hbox{$<$}\mkern-14mu
             \lower0.6ex\hbox{$\sim$}}}
\long\def\symbolfootnote[#1]#2{\begingroup%
\def\thefootnote{\fnsymbol{footnote}}\footnote[#1]{#2}\endgroup}

\journalinfo{The Astrophysical Journal, 
???: ??1--??7, 2011 Month, }
\slugcomment{Received 2011 Dec 00; Accepted 2011 Jan 00}

\shorttitle{XRF 100316D / SN~2010bh and the Nature of Gamma Ray Burst Supernovae}

\shortauthors{Z. Cano, et al.}

\begin{document}

\title{XRF 100316D / SN~2010bh and the Nature of Gamma Ray Burst Supernovae\footnote{Based on observations made with the NASA/ESA Hubble Space Telescope, obtained at the Space Telescope Science Institute, which is operated by the Association of Universities for Research in Astronomy, Inc., under NASA contract NAS 5-26555. These observations are associated with program \# 11709.}}

\author{
Z. Cano\altaffilmark{1}$^{,}$\altaffilmark{$\dagger$},
D. Bersier\altaffilmark{1},
C. Guidorzi\altaffilmark{2}$^{,}$\altaffilmark{1},
S. Kobayashi\altaffilmark{1},
A. J. Levan\altaffilmark{3}, 
N. R. Tanvir\altaffilmark{4}, 
K. Wiersema\altaffilmark{4}, 
P. D'Avanzo\altaffilmark{5}, 
A. S. Fruchter\altaffilmark{6},  
P. Garnavich\altaffilmark{7},
A. Gomboc\altaffilmark{8}$^{,}$\altaffilmark{9}, 
J. Gorosabel\altaffilmark{10}, 
D. Kasen\altaffilmark{11}$^{,}$\altaffilmark{12}, 
D. Kopa\v c\altaffilmark{8}, 
R. Margutti\altaffilmark{5}, 
P. A. Mazzali\altaffilmark{13}$^{,}$\altaffilmark{14}, 
A. Melandri\altaffilmark{5}$^{,}$\altaffilmark{1}, 
C. G. Mundell\altaffilmark{1},
P. E. Nugent\altaffilmark{15},
E. Pian\altaffilmark{14}$^{,}$\altaffilmark{16}, 
R. J. Smith\altaffilmark{1},
I. Steele\altaffilmark{1},
R. A. M. J. Wijers\altaffilmark{17},
and
S. E. Woosley\altaffilmark{11} 
}

\altaffiltext{1}{Astrophysics Research Institute, Liverpool John Moores University, Liverpool, UK.} 
\altaffiltext{2}{Dipartimento di Fisica, Universit\' a di Ferrara, via Saragat 1, I-44100 Ferrara, Italy.}
\altaffiltext{3}{Department of Physics, University of Warwick, Coventry UK.}
\altaffiltext{4}{Department of Physics and Astronomy, University of Leicester, Leicester, UK.}
\altaffiltext{5}{INAF - Osservatorio Astronomico di Brera, via E. Bianchi 46, 23807, Merate, LC, Italy.}
\altaffiltext{6}{Space Telescope Science Institute, Baltimore, MD, USA.}
\altaffiltext{7}{Physics Dept., University of Notre Dame, Notre Dame, IN, USA.}
\altaffiltext{8}{Faculty of Mathematics and Physics, University of Ljubljana, Slovenia.}
\altaffiltext{9}{Centre of Excellence SPACE-SI, A\v sker\v ceva cesta 12, SI-1000 Ljubljana Slovenia.}
\altaffiltext{10}{Instituto de Astrof\' isica de Andaluc\' ia (IAA-CSIC), Granada, Spain.} 
\altaffiltext{11}{University of California, Santa Cruz, CA, USA.}
\altaffiltext{12}{Hubble Fellow.}
\altaffiltext{13}{Max-Planck-Institut f\"ur Astrophysik, Karl-Schwarzschild- Strasse 1, Garching, Germany.}
\altaffiltext{14}{INAF Oss. Astron. Padova, vicolo dellOsservatorio 5, Padova, Italy.}
\altaffiltext{15}{Computational Cosmology Center, Lawrence Berkeley National Laboratory, Berkeley, CA, USA.}
\altaffiltext{16}{Osservatorio Astronomico Di Trieste, Via G.B.Tiepolo, Trieste, Italy.}
\altaffiltext{17}{Astronomical Institute, University of Amsterdam, XH Amsterdam, The Netherlands.}

\altaffiltext{$\dagger$}{zec@astro.livjm.ac.uk}

\begin{abstract}

We present ground-based and \emph{HST} optical and infrared
observations of \emph{Swift} XRF 100316D / SN~2010bh.  It is seen that the optical
light curves of SN~2010bh evolve at a faster rate than the archetype
GRB-SN~1998bw, but at a similar rate to SN~2006aj, a supernova that
was spectroscopically linked with XRF 060218, and at a similar rate to
non-GRB associated type Ic SN~1994I.  We estimate the rest-frame
extinction of this event from our optical data to be
$E(B-V)=0.18\pm0.08$ mag.  We find the $V$-band absolute magnitude of
SN~2010bh to be $M_{V}=-18.62\pm0.08$, which is the faintest peak
$V$-band magnitude observed to-date for a spectroscopically-confirmed GRB-SNe.  When we investigate the origin of
the flux at $\rm{t-t_{o}=0.598}$ days, it is shown that the light is not
synchrotron in origin, but is likely coming from the supernova shock
break-out.  We then use our optical and infrared data to create a
quasi-bolometric light curve of SN~2010bh which we model with a simple
analytical formula.  The results of our modeling imply that SN~2010bh
synthesized a nickel mass of $\rm{M_{Ni}\approx 0.1 M_{\odot}}$, 
ejected  $\rm{M_{ej}\approx 2.2 M_{\odot}}$ and has an explosion
energy of $\rm{E_{k}\approx 1.4 \times 10^{52}}$ erg.  Thus, while SN~2010bh
is an energetic explosion, the amount of nickel created during the
explosion is much less than that of SN~1998bw, and only marginally
more than SN~1994I.  Finally, for a sample $22$ GRB-SNe we check for a correlation
between the stretch factors and luminosity factors in the $R$ band and conclude that no statistically-significant
correlation exists.

\end{abstract}


\section{Introduction}
\label{introduction}

For more than a decade the intriguing connection between long-duration gamma ray bursts (GRBs) and X-ray Flashes (XRFs) with type Ibc, stripped-envelope, core-collapse supernovae (SNe) has been studied in considerable depth.  What is now the archetype gamma ray burst-supernova (GRB-SN), SN~1998bw (e.g. Galama et al. 1998a; Patat et al. 2001) provided the first observational evidence that GRBs are generated during the collapse of a massive star.  Since then several other detections of GRB-SNe have strengthened the link between the collapse of massive stars and GRBs, including: GRB 030329 / SN~2003dh (Hjorth et al. 2003; Stanek et al. 2003; Matheson et al. 2003), GRB 031203 / SN~2003lw (e.g. Malesani et al. 2004; Gal-Yam et al. 2004) and XRF 060218 / SN~2006aj (e.g. Pian et al. 2006; Campana et al. 2006; Ferrero et al. 2006; Mazzali et al. 2006).  In all of the aforementioned cases, spectroscopy of each event revealed unusually broad spectral features, indicative of much higher ejecta velocities that are much higher than those seen in ``typical'' Ic SN events.

However, there exists not just a simple dichotomy between typical type Ic SNe such as SN~1994I (e.g. Richmond et al. 1996; Iwamoto et al. 1994) that have modest peak magnitudes and ejecta moving at velocities anywhere from $8,000-18,000$ km s$^{-1}$ (e.g. Matheson et al. 2001), and GRB-SNe that generally have brighter peak magnitudes and ejecta velocities up to $30,000$ km s$^{-1}$ (e.g. SN~1998bw; Galama et al. 1998a).  Indeed the type Ic SN class is quite heterogeneous, especially when the discoveries of broad-lined SNe that are not accompanied by a GRB trigger are also considered.  Events such as SN~2007gr (Paragi et al. 2010; though see Soderberg et al. 2010a) and SN~2009bb (Stritzinger et al. 2009; Pignata et al. 2011) reveal that relativistic ejecta are present in some non-GRB associated SNe.  Even more intriguingly, Soderberg et al. (2010b) showed 
that SN~2009bb most likely had a prolonged central engine.  The multi-frequency radio observations presented by Soderberg et al. (2010b) are well described by a self-absorbed synchrotron spectrum which they suggest is produced as the shock break-out accelerated electrons in the local circumstellar medium.  

Thus, it seems that the special requisites for GRB-production are very atypical of the general Ic SNe population; indeed it is possible to have an energetic type Ic SN with a bright maximum, broad spectral features, and prolonged central engine activity, without the production of a GRB.  Not suprisingly, the differences between GRB-SNe and local Ic SNe without a GRB-trigger remains elusive, though the in-depth investigation of individual cases provides an excellent opportunity for understanding these enigmatic events.  Previous investigations into individual events has revealed that no two GRB-SNe are the same, and the latest example we present in this paper is no exception.

XRF 100316D was detected by \emph{Swift} on the 16$^{th}$ March, 2010 at 12:44:50 UT.  The redshift was measured to be $z=0.0591 \pm 0.0001$ (Vergani et al. 2010a; Starling et al. 2011; Chornock et al. 2010), and we use this value throughout the paper.  Spectroscopy presented by Chornock et al. (2010) revealed similar behaviour to previous GRB-SNe, where the spectra are 
featureless during the first few days, and with the emergence of broad spectral features after several days.  As for other GRB-SNe, Chornock et al. (2010) reported that there was no evidence for helium in their spectroscopy, indicating that the progenitor is a highly-stripped star.  However, spectroscopy taken by Bufano et al. (2011) revealed a broad He absorption feature at $\rm 10830 \AA$, a spectral line that has only be observed one other time in a GRB-SNe (SN~1998bw, Patat et al. 2001).  Bufano et al. (2011) also detected He in their early spectra at $\rm 5876 \AA$.  These detections of He indicate that the progenitor star still retained at least a small fraction of its He envelope prior to explosion, indicating that SN~2010bh is a type Ibc SNe.

Starling et al. (2011) 
found that the high-energy properties of XRF 100316D / SN 2010bh are quite different from those of other GRB-SNe, but are similar to the high energy properties of XRF 060218 / SN~2006aj.  Starling et al. (2011) showed in the X-ray spectrum at $144$ s $\le \rm{t-t_{o}} \le 737$ s the presence of a soft, hot, black-body component that contributes $\approx 3 \%$ to the total $0.3-10.0$ keV flux (though see also Fan et al. 2011 who claim that both a Cutoff Power-law model or a Cutoff Power-law model + black-body component provide equally acceptable fits).  The presence of the black-body component ($kT = 0.14$ keV; $T \approx 1.6 \times 10^{6}$ K) is similar in temperature to the soft, hot, black-body component seen in the X-ray spectrum of XRF 060218 / SN~2006aj ($kT \approx 0.17$ keV; $T \approx 2.0 \times 10^{6}$ K; Campana et al. 2006).

A detailed discussion of the host galaxy metallicity at the explosion site of SN~2010bh was performed by Levesque et al. (2011).  The authors presented evidence, which was also reported by Chornock et al. (2010) and Starling et al. (2011), that SN~2010bh occurred in a star-forming region of the host galaxy that has a star-formation rate of $\sim 1.7 \rm M_{\odot}$ yr$^{-1}$ (at the site of the supernova) and low metallicity ($Z \le 0.4 Z_{\odot}$; Chornock et al. (2010), Levesque et al. (2011)).

In this paper we present in Section~\ref{sec:obs} optical and infrared photometry of XRF 100316D / SN~2010bh obtained on Faulkes Telescope South (FTS); Gemini South (Gemini-S) and the Hubble Space Telescope (\emph{HST}). In Section~\ref{sec:lc} we use our optical light curves (LCs) to derive the peak optical properties of SN~2010bh, showing that SN~2010bh evolves much quicker than SN~1998bw, and at a rate similar to SN~2006aj.  We also discuss in Section 3 the origin of the flux at $\rm{t-t_{o}=0.598}$ days, concluding that it is not synchrotron but is likely coming from the shock break-out.  In Section~\ref{sec:ebv} we make an estimate of the rest frame extinction, and derive peak absolute magnitudes of SN~2010bh.  In Section~\ref{sec:bol} we use our optical and infrared (IR) photometry, along with our estimation of the host extinction, to construct the bolometric light curve, and use an analytical model to extract physical parameters of the SN.  
In Section~\ref{sec:discuss} we discuss SN~2010bh in relation to SN~2006aj, as well as other GRB-SNe and local type Ic SNe that were not accompanied by a GRB.  In the discussion we investigate the light curve properties of most of the spectroscopically and photometrically-linked GRB-SNe.  By making simple assumptions, we obtain host- and afterglow-subtracted ``supernova'' light curves, and compare them with SN~1998bw, deriving stretch ($s$) and luminosity ($k$) factors of the GRB-SNe in relation to SN~1998bw.  When then check for a correlation between the stretch and luminosity factors, concluding that no statistically-significant correlation exists.

Throughout the paper observer-frame times are used unless specified otherwise in the text.  Foreground reddening of $E(B-V) = 0.117$ mag has been corrected for using the dust maps of Schlegel, Finkbeiner \& Davis (1998).  We adopt a flat $\Lambda$CDM cosmology with $H_{0}=71$ km s$^{-1}$ Mpc$^{-1}$, $\rm{\Omega_{M}} = 0.27$, and $\rm{\Omega_{\Lambda}=1-\Omega_{M}} =0.73$.  For this cosmology a redshift of $z=0.0591$ corresponds to a luminosity distance of $\rm{d_{L}} = 261$ Mpc and a distance modulus of $37.08$ mag.

\section{Observations \& Photometry}
\label{sec:obs}

\begin{figure*}
\centering
\includegraphics[height=7in,width=5in, angle=-90]{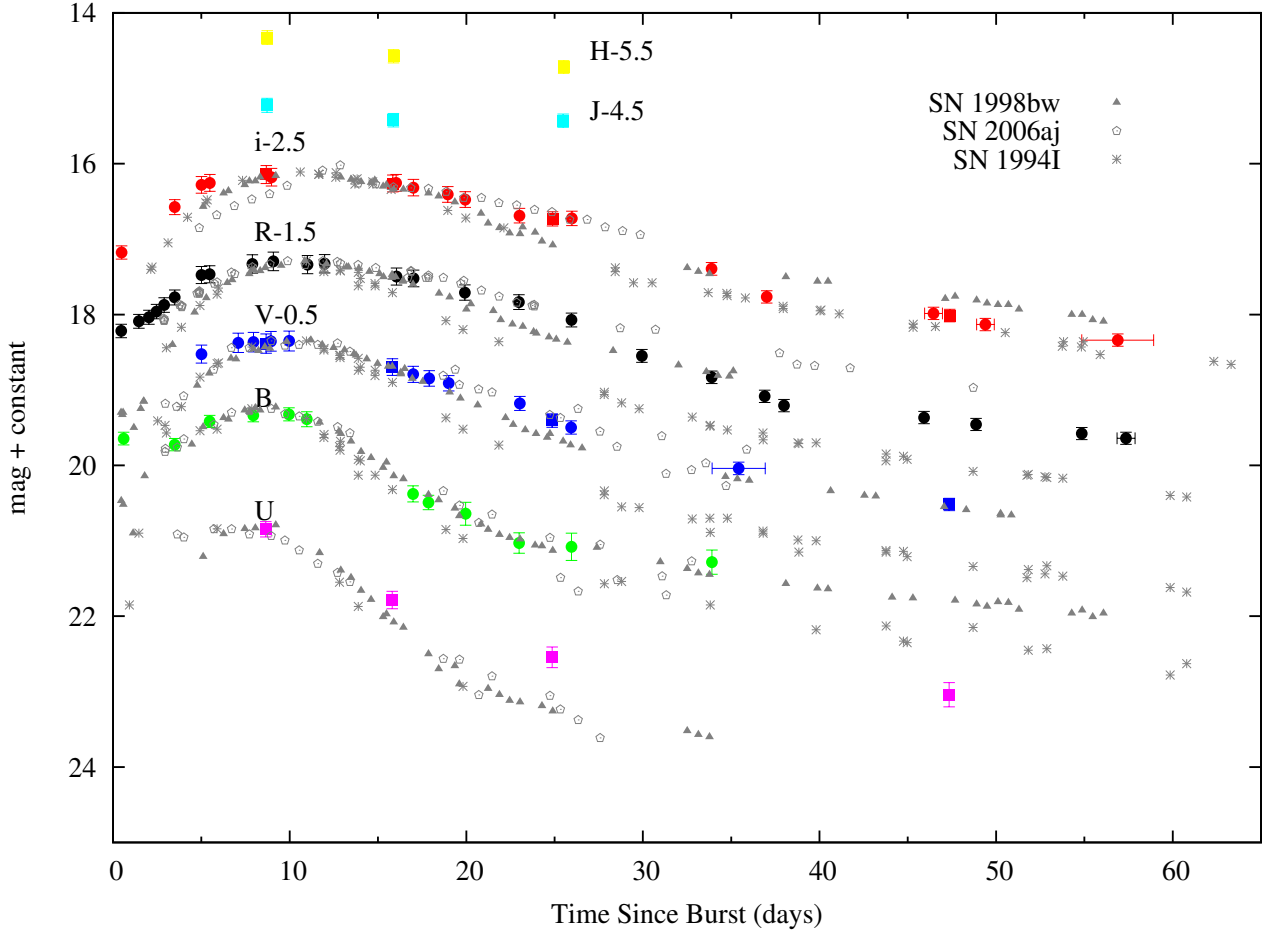}
\caption{$UBVR_{c}iJH$ light curves of XRF 100316D /SN 2010bh (solid points (ground-based data) and filled squares (\emph{HST})).  Plotted for comparison are the light curves of two known GRB-SNe: SN~1998bw (Galama et al. 1998a; Patat et al. 2001; Sollerman et al. 2000) and SN~2006aj (Campana et al. 2006; Sollerman et al. 2006; Ferrero et al. 2006), as well as a local SN that is not associated with a GRB: SN~1994I (Richmond et al. 1996).  All of the SN~2010bh light curves have been corrected for foreground and host extinction.  The light curves of the comparison SNe have been shifted by different amounts in each filter to match in peak brightness.  No other changes have been made to the light curves apart from the light curves of SN~1998bw which have been stretched in each filter by a factor of $\sim 0.6$ (see Table~\ref{table:2010bh_stretch})
and the early-time data ($\rm{t-t_{o}< 3}$ days) for SN~2006aj, for which the light was dominated by flux coming from the shock break-out, was not included to avoid confusion between different filters. }
\label{fig:GRB100316D_LCs}
\end{figure*}

\subsection{Data Acquisition \& Calibration}

\textbf{\textit{Faulkes Telescope South \& Gemini-South}}\ \\

We obtained forty-one epochs of photometric data between March and May 2010 using the $2$m Faulkes Telescope South (FTS).  Observations were made using the $4.6'$ x $4.6'$ field of view Merope Camera in filters $BVR_{c}i$.  A $42^{nd}$ and final epoch of observations were made in December 2010 in all filters, which were used as template/reference images for image subtraction (see Section 2.2).  Images were also taken of Landolt photometric standard regions (Landolt 1992) in $BVR_{c}i$ on the $2^{nd}$ of May, 2010 using the Spectral Camera, which with its wider field of view ($10.5'$ x $10.5'$), allowed us to calibrate over 140 secondary standards in the field of XRF 100316D.  Subsequent observations made on Gemini-South and \emph{HST} are calibrated to these secondary standards.

Five epochs of photometric data were obtained on Gemini-South \footnote{The data presented here supersedes those published in the GCN Circulars: GCNs 10513 (Vergani et al. 2010b); 10523 (Levan et al. 2010); 10525 (Wiersema et al. 2010).} (Gemini-S) in filters $griz$, the first starting only $\approx 0.5$d after the initial burst.  A final epoch was obtained on the $28^{th}$ of January, 2011 in all filters, which were used as template images for image subtraction (see Section 2.2).  

\ \\

\noindent\textbf{\textit{Hubble Space Telescope}}\ \\

We obtained five epochs of photometric data with $HST$, using the Wide Field Camera 3 (WFC3).  The first three epochs of data were obtained in filters $F336W$, $F555W$, $F814W$, $F125W$ and $F160W$, while the fourth epoch yielded images in filters $F336W$, $F555W$, $F814W$, and the fifth epoch in filters $F555W$ and $F814W$.  
We find for the optical transient (OT) associated with XRF 100316D a position of: $07^{h}10^{m}30.54^{s}(\pm0.02)^{s}$, $-56^{d}15^{'}19.80^{''}(\pm0.10)^{''}$.  

Aperture photometry was performed on the \emph{HST} images using standard routines in IRAF\footnote{IRAF is distributed by the National Optical Astronomy Observatory, which is operated by the Association of Universities for Research in Astronomy, Inc., under cooperative agreement with the National Science Foundation.}.  A small aperture was used, and an aperture correction for an isolated star was computed and applied.  The aperture-corrected, instrumental magnitudes were then calibrated via standard star photometry to secondary standards in the field of XRF 100316D.  Images taken in filters $F555W$ and $F814W$ are calibrated respectively to filters $V$ and $i$ using a zero-point and a colour term.  The remaining filters are calibrated to AB magnitudes using the appropriate AB zero-point for each filter that are listed in the WFC3 instrument and data-handbook\footnote{http://www.stsci.edu/hst/wfc3}.  The \emph{HST} magnitudes, which have been corrected for foreground extinction, are listed in Table \ref{table:2010bh_HST}.  The quoted errors, which are added in quadrature, are derived from the uncertainties associated with the photometry and calibration.

\begin{table}
\begin{center}
\caption{\textit{HST} Photometry of XRF 100316D / SN~2010bh 
\label{table:2010bh_HST}}
\begin{tabular}{ccccc}
\hline
 Filter & $\rm{t-t_{o}}$ (days)  & Mag & $\sigma$ (mag)  & Calibrated to\\
\hline
$F336W$	&	8.650	&	21.86	&	0.06	&	AB	\\
$F336W$	&	15.810	&	22.80	&	0.08	&	AB	\\
$F336W$	&	24.860	&	23.56	&	0.11	&	AB	\\
$F336W$	&	47.340	&	24.05	&	0.14	&	AB	\\
$V$	&	8.663	&	19.50	&	0.01	&	Vega	\\
$V$	&	15.822	&	19.81	&	0.01	&	Vega	\\
$V$	&	24.869	&	20.52	&	0.02	&	Vega	\\
$V$	&	47.352	&	21.63	&	0.03	&	Vega	\\
$V$	&	137.490	&	23.07	&	0.08	&	Vega	\\
$i$	&	8.675	&	19.00	&	0.01	&	AB	\\
$i$	&	15.844	&	19.12	&	0.01	&	AB	\\
$i$	&	24.881	&	19.59	&	0.01	&	AB	\\
$i$	&	47.370	&	20.87	&	0.02	&	AB	\\
$i$	&	137.518	&	22.70	&	0.06	&	AB	\\
$F125W$	&	8.720	&	19.89	&	0.07	&	AB	\\
$F125W$	&	15.870	&	20.09	&	0.07	&	AB	\\
$F125W$	&	25.480	&	20.10	&	0.07	&	AB	\\
$F160W$	&	8.730	&	19.93	&	0.08	&	AB	\\
$F160W$	&	15.880	&	20.18	&	0.08	&	AB	\\
$F160W$	&	25.520	&	20.32	&	0.07	&	AB	\\
\hline
\end{tabular}

\medskip
All magnitudes have been corrected for foreground extinction.
\end{center}
\end{table}

\subsection{Image Subtraction}

Using ISIS (Alard 2000), we performed image subtraction on the FTS and Gemini-S images, using the last epoch of images taken in each filter and on each telescope as the respective templates.  Subtracting the template from the early FTS images gave a clear detection of the optical transient (OT) and an accurate position, and the same was also observed for the Gemini-S images.

Aperture photometry was then performed on the subtracted images using a small aperture and an aperture correction was computed from the pre-subtracted images and applied.  The instrumental magnitudes obtained from the FTS subtracted images are calibrated to $BVR_{c}i$.  The Gemini-S subtracted images taken in filters $griz$ are calibrated to $BR_{c}iz$ respectively.  Images in filters $gr$ are calibrated using a zero-point and a colour term, while images taken in $i$ are calibrated using only a zero-point.  Magnitudes of the secondary standards in $z$ are computed using transformation equations from Jester et al. (2005), and the Gemini-S data are calibrated to these using a zero-point.  All of the image-subtracted, foreground-corrected magnitudes obtained from the FTS and Gemini-S data are displayed in Table \ref{table:2010bh_ground}, where the quoted errors (added in quadrature) are determined from the photometry and calibration.

\begin{table*}
\begin{center}
\scriptsize
\caption{Ground-based Photometry of XRF 100316D / SN~2010bh 
\label{table:2010bh_ground}}
\begin{tabular}{c c c c c c c c c c c c}
\hline
 Filter & $\rm{t-t_{o}}$ (days)  & Mag & $\sigma$ (mag)  & Calibrated to & Telescope$^{a}$ &  Filter & $\rm{t-t_{o}}$ (days)  & Mag & $\sigma$ (mag)  & Calibrated to & Telescope$^{a}$\\
\hline
$B$ &  0.598 & 20.45 & 0.03 & Vega & GS  & $R_{c}$ & 11.015 & 19.33 & 0.02 & Vega & FTS  \\     
$B$ &  3.489 & 20.53 & 0.01 & Vega & GS  & $R_{c}$ & 11.985 & 19.32 & 0.04 & Vega & FTS  \\
$B$ &  5.467 & 20.22 & 0.01 & Vega & GS  & $R_{c}$ & 16.045 & 19.49 & 0.04 & Vega & FTS  \\
$B$ &  7.061 & 20.25 & 0.05 & Vega & FTS & $R_{c}$ & 17.023 & 19.52 & 0.02 & Vega & FTS  \\
$B$ &  7.940 & 20.14 & 0.03 & Vega & FTS & $R_{c}$ & 19.922 & 19.71 & 0.02 & Vega & FTS  \\
$B$ &  9.956 & 20.13 & 0.03 & Vega & FTS & $R_{c}$ & 22.978 & 19.83 & 0.02 & Vega & FTS  \\
$B$ & 10.965 & 20.19 & 0.06 & Vega & FTS & $R_{c}$ & 25.956 & 20.07 & 0.03 & Vega & FTS  \\
$B$ & 16.987 & 21.18 & 0.07 & Vega & FTS & $R_{c}$ & 29.941 & 20.54 & 0.04 & Vega & FTS  \\
$B$ & 17.863 & 21.30 & 0.05 & Vega & FTS & $R_{c}$ & 33.901 & 20.83 & 0.04 & Vega & FTS  \\
$B$ & 19.958 & 21.44 & 0.13 & Vega & FTS & $R_{c}$ & 36.901 & 21.08 & 0.07 & Vega & FTS  \\
$B$ & 22.994 & 21.83 & 0.11 & Vega & FTS & $R_{c}$ & 37.988 & 21.20 & 0.13 & Vega & FTS  \\
$B$ & 25.955 & 21.88 & 0.16 & Vega & FTS & $R_{c}$ & 45.921 & 21.36 & 0.08 & Vega & FTS  \\
$B$ & 30.974 & 22.27 & 0.10 & Vega & FTS & $R_{c}$ & 48.867 & 21.45 & 0.07 & Vega & FTS  \\
$B$ & 33.910 & 22.09 & 0.14 & Vega & FTS & $R_{c}$ & 54.853 & 21.57 & 0.07 & Vega & FTS  \\
$V$ &  5.007 & 19.64 & 0.02 & Vega & FTS & $R_{c}$ & 57.358 & 21.64 & 0.07 & Vega & FTS  \\
$V$ &  7.085 & 19.49 & 0.03 & Vega & FTS & $i$ &  0.479 & 20.04 & 0.01 & AB & GS  \\    
$V$ &  7.956 & 19.48 & 0.02 & Vega & FTS & $i$ &  3.497 & 19.44 & 0.01 & AB & GS  \\    
$V$ &  8.937 & 19.47 & 0.02 & Vega & FTS & $i$ &  5.002 & 19.14 & 0.02 & AB & FTS  \\   
$V$ &  9.970 & 19.47 & 0.03 & Vega & FTS & $i$ &  5.475 & 19.12 & 0.01 & AB & GS  \\    
$V$ & 16.995 & 19.91 & 0.04 & Vega & FTS & $i$ &  8.972 & 19.04 & 0.01 & AB & FTS  \\   
$V$ & 17.899 & 19.96 & 0.02 & Vega & FTS & $i$ & 16.028 & 19.12 & 0.03 & AB & FTS  \\  
$V$ & 19.009 & 20.03 & 0.07 & Vega & FTS & $i$ & 17.005 & 19.18 & 0.02 & AB & FTS  \\  
$V$ & 23.031 & 20.29 & 0.04 & Vega & FTS & $i$ & 18.958 & 19.27 & 0.05 & AB & FTS  \\  
$V$ & 25.939 & 20.61 & 0.04 & Vega & FTS & $i$ & 19.940 & 19.34 & 0.02 & AB & FTS  \\  
$V$ & 35.423 & 21.16 & 0.06 & Vega & FTS & $i$ & 23.013 & 19.55 & 0.03 & AB & FTS  \\  
$R_{c}$ &  0.466 & 20.21 & 0.01 & Vega & GS  & $i$ & 25.977 & 19.59 & 0.05 & AB & FTS  \\  
$R_{c}$ &  1.467 & 20.09 & 0.01 & Vega & GS  & $i$ & 33.892 & 20.26 & 0.08 & AB & FTS  \\  
$R_{c}$ &  2.028 & 20.03 & 0.03 & Vega & FTS & $i$ & 37.014 & 20.63 & 0.08 & AB & FTS  \\  
$R_{c}$ &  2.462 & 19.96 & 0.01 & Vega & GS  & $i$ & 46.456 & 20.85 & 0.07 & AB & FTS  \\  
$R_{c}$ &  2.891 & 19.87 & 0.02 & Vega & FTS & $i$ & 49.393 & 20.99 & 0.06 & AB & FTS  \\  
$R_{c}$ &  3.493 & 19.77 & 0.01 & Vega & GS  & $i$ & 56.887 & 21.20 & 0.09 & AB & FTS  \\  
$R_{c}$ &  5.003 & 19.47 & 0.02 & Vega & FTS & $z$ &  0.470 & 19.91 & 0.11 & AB & GS  \\    
$R_{c}$ &  5.471 & 19.46 & 0.01 & Vega & GS  & $z$ &  3.490 & 19.40 & 0.01 & AB & GS  \\    
$R_{c}$ &  7.891 & 19.33 & 0.02 & Vega & FTS & $z$ &  5.470 & 19.15 & 0.01 & AB & GS  \\    
$R_{c}$ &  9.072 & 19.29 & 0.02 & Vega & FTS &     &       &       &      &    &  \\
\hline
\end{tabular}

\medskip
$^{a}$Telescope key: FTS: $2$m Faulkes Telescope South; GS: $8.1$m Gemini-South Telescope.
All magnitudes have been corrected for foreground extinction.
\end{center}
\end{table*}

\subsection{Host Photometry}

As already noted by Starling et al (2011), the host galaxy system of XRF 100316D / SN~2010bh has a highly disturbed morphology, with a bright central region, a possible spiral arm and a number of bright knots (see Figure 8 in Starling et al. 2011).  SN~2010bh is located on top of a bright knot close to the bright central region, which is perhaps the galactic nucleus.  In our final epoch of Gemini-S images, the host-complex appears as a blended, elliptical object due to the lower resolution of Gemini-S relative to \emph{HST}.  

We have performed photometry on the final Gemini-S images using SExtractor (Bertin \& Arnouts 1996).  We obtained (1) corrected isophotal magnitudes (MAG$\_$ISOCOR), and (2) Kron-like elliptical aperture magnitudes (MAG$\_$AUTO), with the best of these values being used (MAG$\_$BEST).  The host magnitudes in filters $griz$ have been calibrated to secondary standards in the field of XRF 100316D, where the magnitudes of the secondary standards in filters $grz$ are computed using transformation equations from Jester et al. (2005), and the host photometry are calibrated to these using a zero-point.    The Gemini-S photometry, which have been corrected for foreground extinction, are listed in Table \ref{table:2010bh_host}, where the quoted errors of the Gemini-S magnitudes are added in quadrature and are derived from the photometry and calibration.

\section{Optical \& Infrared Light Curves and colour curves}
\label{sec:lc}

\subsection{Light Curves}

The image-subtracted light curves of XRF 100316D / SN2010bh, which have been corrected for foreground and host extinction (see Section 4), are displayed in Figure \ref{fig:GRB100316D_LCs}.  Plotted for comparison are the multi-wavelength light curves of two known GRB-SNe: SN~1998bw and SN~2006aj, as well as SN~1994I which is a local type Ic SN that was not associated with a GRB.  All of the comparison light curves have been shifted in magnitude to match SN~2010bh around maximum light.  No other changes have been made to the light curves apart from the light curves of SN~1998bw, which have been stretched in each filter by the values listed in Table \ref{table:2010bh_stretch} (see Section 3.5), and the early-time data ($\rm{t-t_{o}< 3}$ days) for SN~2006aj, for which the light was dominated by flux coming from the very bright shock break-out, was not included to avoid confusion between different filters.  The stretch-factors have been applied to SN~1998bw for clarity. 

The peak times and magnitudes in each filter are estimated from low-order polynomial fits.  We have determined that SN~2010bh peaked at $\rm{t-t_{o}= 8.02 \pm 0.12}$ days in the $B$ filter, and at subsequently later times in the redder filters.  The quoted errors are statistical and are determined from the distribution of peak times determined from the different order polynomials.  The behaviour of the LC peaking at later times in redder filters is typical behaviour for a core-collapse SNe, and SN~2010bh appears no different in this respect.  The main photometric parameters of XRF 100316D / SN~2010bh in filters $BVR_{c}i$ are presented in Table \ref{table:2010bh_phot_params}.

Comparing the optical light curves of SN~2010bh with those of SN~2006aj, it appears they evolve at roughly the same rate in the $B$ filter.  However it appears that the $U$-band light curve of SN~2006aj evolves faster than SN~2010bh, while in the redder $V$, $R_{c}$ and $i$ filters it evolves more slowly.  It is also worth noting that the $U$-band measurement of SN~2010bh at $\rm{t-t_{o}=24.8}$ days is $\approx 0.5$ mag brighter than the shifted LC of SN~2006aj, and perhaps even more at $\rm{t-t_{o}=47.3}$ days.  As the fluxes are measured in a fixed, and albeit small, aperture, the flux we are detecting is from both the SN and the underlying host galaxy.  When performing photometry on the third and fourth epoch $U$-band images, it appears that the background/host flux is becoming dominant and affecting the overall shape of the LC, (i.e. making it appear that it is fading \emph{slower} than is expected).

\begin{table}
\begin{center}
\begin{minipage}{80mm}
\centering
\caption{Gemini-S Host Photometry of XRF 100316D / SN~2010bh 
\label{table:2010bh_host}}
\begin{tabular}{ccccc}
\hline
 Filter &  Mag & $\sigma$ (mag)  & Telescope\\
\hline
$g$ & 17.40 & 0.08 & Gemini-S \\
$r$ & 17.14 & 0.07 & Gemini-S \\
$i$ & 17.07 & 0.06 & Gemini-S \\
$z$ & 17.03 & 0.06 & Gemini-S \\
\hline
\end{tabular}
\medskip

All magnitudes have been corrected for foreground extinction.
\end{minipage}
\end{center}
\end{table}

\begin{table*}
\begin{center}
\caption{Main Photometric parameters of XRF 100316D / SN~2010bh} 
\label{table:2010bh_phot_params}
\begin{tabular}{rcccc}
\hline
  & $B$ & $V$ & $R_{c}$ & $i$\\
\hline
peak apparent magnitude & $20.13 \pm 0.08$ & $19.47 \pm 0.08$ & $19.29 \pm 0.08$ & $19.05 \pm 0.08$ \\
peak absolute magnitude$^{a}$ & $-18.27 \pm 0.08$ & $-18.62 \pm 0.08$ & $-18.60 \pm 0.08$ & $-18.62 \pm 0.08$ \\
$T_{\rm peak}$ (days) & $8.02\pm 0.12$ & $8.86\pm 0.07$ & $11.18\pm 0.05$ & $11.21\pm 0.22$ \\
$A_{\nu,\rm fore+host}$ (mag) & $1.32 \pm 0.08$ & $1.01 \pm 0.08$ & $0.81 \pm 0.08$ & $0.59 \pm 0.08$ \\
$\Delta m_{15}$ (mag) & $1.80\pm 0.14$ & $0.95\pm 0.10$ & $0.90\pm 0.05$ & $0.77\pm 0.20$ \\
\hline
\end{tabular}
\medskip

$^{a}$ Using a distance modulus of $\mu = 37.08$ mag (for $H_{o}=71$km/s/Mpc, $\Omega_{M} = 0.27$ and $\Omega_{\Lambda}=1-\Omega_{M} =0.73$).\\
Apparent magnitudes are not corrected for foreground or host extinction.  Absolute magnitudes \emph{are} extinction corrected (foreground \& host).
\end{center}
\end{table*}

\subsection{Colour Curves}

\begin{figure}
\centering
\includegraphics[scale=0.36, angle=-90]{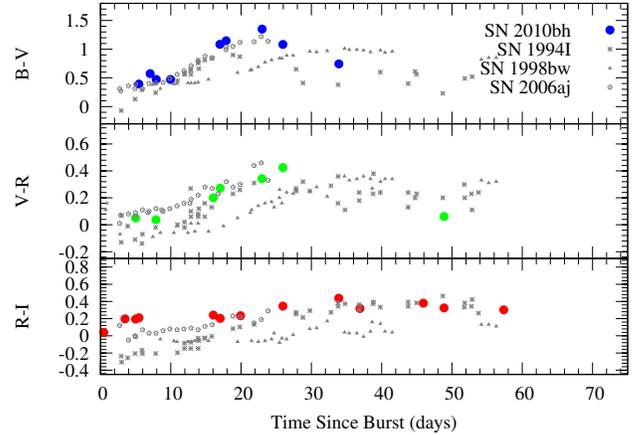}
\caption{Colour curves of XRF 100316D/SN~2010bh, which are corrected for foreground and host extinction.  Plotted for comparison are the colour curves of known GRB-SNe: 1998bw and 2006aj as well as local type Ic SN~1994I that was not accompanied by a GRB trigger.  No other changes have been made to the colour curves apart from those of SN~1998bw, which have been stretched by $s=0.63$ in $B-V$ and $V-R$ and by $s=0.61$ in $R_{c}-i$ (see Section 3.5).  The stretch factors, which are the average values of $s$ determined in the individual filters, have been applied so as to present a consistent analysis of the properties between SN~2010bh and SN~1998bw.}
\label{fig:GRB100316D_colour}
\end{figure}

The colour curves of SN~2010bh are plotted in Figure \ref{fig:GRB100316D_colour}, and have been corrected for foreground and host extinction (see Section 4).  Plotted for comparison are the colour curves of GRB-SNe 1998bw and 2006aj, as well as local SN~1994I, all of which have been corrected for foreground and host extinction using the values quoted in the literature: SN~1994I: $E(B-V)_{\rm total}=0.45$ mag (Richmond et al. 1996; Iwamoto et al. 1996; though see Sauer et al. 2006 who find $E(B-V)=0.30$ mag from spectral modelling); SN~1998bw: $E(B-V)_{\rm total}=0.07$ mag (Patat et al. 2001); SN~2006aj: $E(B-V)_{\rm total}=0.14$ mag (Sollerman et al. 2006; though see Campana et al. 2006 who find $E(B-V)_{\rm host}=0.20$ mag from their \emph{UVOT} obsverations.).  No other changes have been made to the colour curves apart from those of SN~1998bw, which have been stretched by $s=0.63$ in $B-V$ and $V-R_{c}$ and by $s=0.61$ in $R_{c}-i$ (see Table \ref{table:2010bh_stretch}).  The stretch factors, which are the average values of $s$ determined in the individual filters (i.e. $s=0.62$,$0.64$ for respective filters $B$ and $V$; thus the average value is $s=0.63$), have been applied so as to present a consistent analysis of the properties between SN~2010bh and SN~1998bw (i.e. as the stretch factors were also applied to the LCs of SN~1998bw in Figure \ref{fig:GRB100316D_LCs}).

Starting before maximum light, all of the colours of SN~2010bh become redder over time, with similar behaviour seen in the comparison SNe.  
%
Additionally, SN~2010bh appears to be redder than SN~1998bw and SN~1994I in the pre-maximum phases, while in the post-maximum phase SN~2010bh appears to have colours similar to those seen in SN~1994I.

\subsection{$\Delta \lowercase{m_{15}}$}

We have determined the $\Delta m_{15}$ parameter (i.e. the amount that the light curves fades by from peak to fifteen days later) of the $BVR_{c}i$ light curves of XRF 100316D / SN~2010bh.  We have used low-order polynomials to determine the peak times of each light curve, and in turn the $\Delta m_{15}$ parameter.  It is seen that the light curves decay more slowly in the redder filters, with the $\Delta m_{15}$ parameter in filters $BVR_{c}i$, respectively, being $1.80\pm0.14$, $0.95\pm0.10$, $0.90\pm0.05$ and $0.77\pm0.20$ (which are also listed in Table \ref{table:2010bh_stretch}).  The uncertainties in the $\Delta m_{15}$ parameters are derived from the uncertainties in determining the peak times from the photometry, which, for example, in the $i$ band are less certain due to the sparseness of photometry around and just after the peak, as well as from the distribution of the peak times determined from fitting different order polynomials.

The $V$- and $R$-band values of $\Delta m_{15}$ measured for SN~2010bh are somewhat larger (i.e. fade faster) than the average values measured for a sample of local Ibc SNe observed in the $V$ band and $R$ band by Drout et al. (2011) (their Table 4).  It is also seen that SN~2010bh fades faster, on average, than Ia SNe in the $B$-band (e.g. Phillips 1993; Riess et al. 1998).

\subsection{Shock Break-out or optical afterglow?}

\begin{figure}
\centering
\includegraphics[scale=0.35, angle=-90]{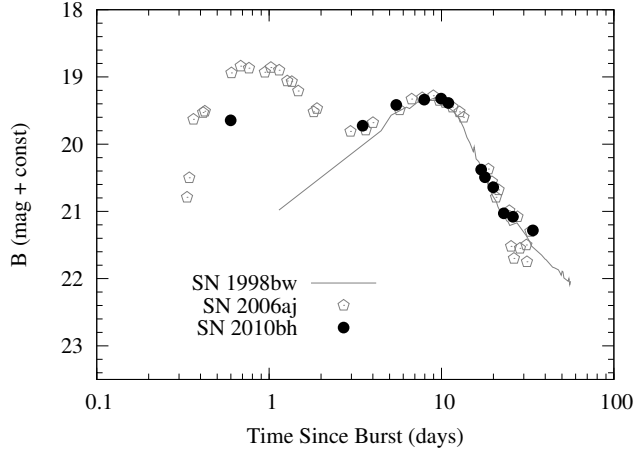}
\caption{$B$-band light curve of XRF 100316D / SN~2010bh (solid points), which has been corrected for foreground and host extinction.  Plotted for comparison are the $B$-band light curves of two known GRB-SNe: 1998bw and 2006aj.  The light curves of the comparison SNe have been shifted by different amounts in each filter to match in peak brightness.  No other changes have been made to the light curves apart for the light curves of SN~1998bw which have been stretched by a factor $s=0.62$ (see Table \ref{table:2010bh_stretch}).  It is seen that SN~2010bh did not increase linearly in brightness from first detection like SN~1998bw, and at $\rm{t-t_{o}=0.59}$ days is of comparable brightness as the shifted light curve of SN~2006aj.}
\label{fig:GRB100316D_B}
\end{figure}

\begin{figure}
\centering
\includegraphics[scale=0.35, angle=-90]{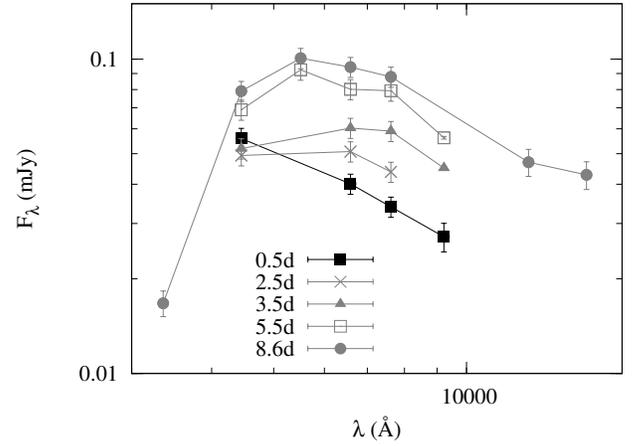}
\caption{Spectral Energy Distributions (SEDs) of XRF 100316D / SN~2010bh over the first $8.6$ days.  All photometry has been corrected for foreground and host extinction.  The first five epochs ($\rm{t-t_{o}} \approx 0.5$, $1.5$, $2.5$, $3.5$ and $5.5$ days) are from the Gemini-S detections, while the SED at $t-t_{o}=8.5$ days is from the first \emph{HST} data.  The power-law index (i.e. $F_{\nu} \propto \nu^{\beta}$) of the first SED is $\beta = +0.94 \pm 0.05$; $\chi^{2}/\rm{d.o.f}=2.17/2$ (best-fitting line not plotted).  The slope of the spectrum is steeper than that expected for light emitted as synchrotron radiation, where, when neglecting effects due to self-absorption, a maximum value of $\beta = +1/3$ is allowed in synchrotron theory (e.g. Sari et al. 1998).  Thus, the origin of light is not expected to be synchrotron but rather is coming from the shock break-out from the stellar surface and dense stellar wind. }
\label{fig:GRB100316D_SED}
\end{figure}

The break-out of the blast/shock-wave and/or emission coming from the shock-heated stellar envelope, has been detected in both type Ibc and type II SNe, including:  type IIp SN~1987A (Early peaks in optical LCs: Hamuy et al. 1988, and PCyg-like feature around $\rm 1500 \AA$ in UV light echo off of a dust cloud $\sim 300$ pc from the SN: Gilmozzi \& Panagia 1999); type IIb SN~1993J (Early peak in optical light curves: Wheeler et al. 1993, Schmidt et al. 1993, Richmond et al. 1994, Lewis et al. 1994); type Ibc SN~1999ex (early optical peaks: Stritzinger et al. 2002); type IIP SNe SNLS-04D2dc \& SNLS-06D1jd (UV-flash: Gezari et al. 2008); type IIP GALEX supernova SNLS-04D2dc (UV-flash: Schawinski et al. 2008); type IIn PTF 09UJ (UV-flash: Ofek et al. 2010); and type IIP SN~2010aq (early UV \& optical peaks: Gezari et al. 2010).  

One SN in particular has been the subject of much interest and debate when intrepreting the high energy properties of the event.  SN~2008D (Modjaz et al. 2008; Soderberg et al. 2008; Mazzali et al. 2008; Malesani et al. 2009) occurred in nearby galaxy NGC 2770 while \emph{Swift} was observing SN~2007uy within the same galaxy.  The resultant high-energy emission detected by \emph{Swift} was interpreted differently in the literature, with the hot, black-body X-ray spectrum \& early peak in optical LC being intrepreted by Soderberg et al. 2008 as being due to the shock break-out, while Mazzali et al. (2008) suggests that the emission is due to a ``choked'' jet.  Interestingly, a recent paper by Van der Horst et al. (2011) has perhaps put the debate to rest.  Using observations of the radio emission of SN~2008D during the first year after the explosion, the authors showed that there was no evidence of a relativistic jet contributing to the observed radio flux, thus strongly suggesting that the high-energy emission was due to the shock break-out and not to a GRB- or XRF-like event.

Emission coming from the shock-heat, expanding stellar evelope as also been detected in GRB/XRF-SNe, though the topic is still contentiously debated in the literature.  For XRF 060218, the hot, black-body X-ray spectrum \& early peak in UV and optical LCs was intrepreted by Campana et al. 2006 (C06 here-on), and later Waxman et al. (2007), as being due to the shock break-out.  However, the emission was intrepreted by Ghisellini et al. (2006) as being due to a relativistic jet.  

Figure \ref{fig:GRB100316D_B} shows the $B$-band LC of SN~2010bh, which has been corrected for foreground and host extinction.  Plotted for comparison are the light curves of two GRB-SNe: 1998bw and 2006aj, which have been shifted to match SN~2010bh in peak brightness, and the LC of SN~1998bw has also been stretched by a factor $s=0.62$ (see Table \ref{table:2010bh_stretch}).  

Upon inspection, similarities between SN~2010bh and SN~2006aj are seen in the early-time LC, where 
a bright component  is detected at $\rm{t-t_{o} = 0.598}$ days.   The magnitude at this epoch is similar to that at $\rm{t-t_{o} = 3.489}$ days, indicating that SN~2010bh did not increase linearly in brightness from detection, as was seen for SN~1998bw.  In the case of SN~2006aj, early peaks in the UV and optical LCs ($\mathrm{t-t_{o}}\le 10^{4}$ s) were explained by C06 as light coming from the low energy tail of the thermal X-ray emission (see below) produced by the radiation shock driven into the stellar wind. At later times, the optical and UV emission is attributed to the expanding envelope of the progenitor star that is heated as the shock-wave passes through it.  Initially the envelope is opaque due to the dense stellar wind, but as the star and wind expands, the photosphere propogates inward exposing the shocked stellar material.  For SN~2010bh, though we are not able to ascertain the exact shape of the early $B$-band light curve due to the paucity of our early-time observations, it is not unreasonable to envision a scenario similar to that observed for SN~2006aj in which the brightness of the flux at this early epoch is due to flux coming from the shock-heated, expanding stellar envelope.

In addition to the bright $B$-band detection, inspection of the spectral energy distribution (SED) at this epoch, which has been corrected for foreground and host extinction, gives additional clues to the origin of the flux.  At the first epoch the SED is very blue 
and does not resemble the shape of the SEDs at later times (see Fig~\ref{fig:GRB100316D_SED}).

To determine whether the flux at this time is synchrotron in origin, the optical SED was fit with a power-law (i.e. $F_{\nu} \propto \nu^{\beta}$), for which we find $\beta = +0.94 \pm 0.05$ ($\chi^{2}/\rm{d.o.f}=2.17/2$).  The slope of the spectrum is harder than that expected for light emitted as synchrotron radiation, where, when neglecting effects due to self-absorption (see below), a maximum value of $\beta = +1/3$ is allowed (e.g. Sari et al. 1998).  Thus, the shape of the SED at this early epoch is consistent with emission coming from the shock-heated stellar envelope.

Additionally, we have constructed SEDs for SN~2006aj at $\rm{t-t_{o}}=0.50$ days and 0.75 days and fit them with a single power-law.  Quite interestingly, at $\rm{t-t_{o}}=0.50$ days we find $\beta = +0.97 \pm 0.05$ ($\chi^{2}/\rm{d.o.f}=2.13/2$), while at $\rm {t-t_{o}}=0.75$ days we find $\beta = +0.94 \pm 0.03$ ($\chi^{2}/\rm{d.o.f}=2.03/2$).  These values for the spectral index are akin to that found for SN~2010bh at a similar epoch, and are both harder than that expected for synchrotron radiation.

The assumption that the optical flux is above the self-absorption frequency ($\nu_{a}$) cannot be proved with our current set of observations, however it has been seen in other GRB events that the synchrotron self-absorption frequency at typical observing epochs of about 1 day after the GRB is expected to be of order $\sim 10^{9}-10^{10}$ Hz (i.e. at radio frequencies).  In the literature there are several events where there has been enough multi-wavelength observations to determine the self-absorption frequency, including: GRB 970508: $\nu_{a}\sim 3 \times 10^{9}$ Hz (Granot et al. 1999); $\nu_{a}\sim 2.5 \times 10^{9}$ Hz (Galama et al. 1998); GRB 980329: $\nu_{a}\sim 13 \times 10^{9}$ Hz (Taylor et al. 1998); GRB 991208 $\nu_{a}\sim 4-11\times 10^{9}$ Hz (Galama et al. 2000).  As the frequency range of the optical observations lie between $\sim 3-10 \times 10^{14}$ Hz (i.e. $\sim 10^{4}-10^{5}$ times higher frequency), it is unlikely that the optical observations suffer from self-absorption effects.

Further evidence that the light at $\rm{t-t_{o}} = 0.598$ days is coming from the shock-heated stellar envelope is the analysis performed by Starling et al. (2011) on the X-ray spectrum at $144$ s $ \le \rm{t-t_{o}} \le 737$ s.  The authors found evidence for a soft, hot, black-body component that contributes $\approx 3 \%$ to the total $0.3-10.0$ keV flux (though see also Fan et al. 2011 
for a different interpretation).
With this in mind, the presence of the black-body component, with $kT = 0.14$ keV ($T \approx 1.6 \times 10^{6}$ K) is similar in temperature to the soft, hot, black-body component seen in the X-ray spectrum of XRF 060218 / SN~2006aj ($kT \approx 0.17$ keV; $T \approx 2.0 \times 10^{6}$ K; C06).  

According to C06, the thermal component is key to understanding XRF 060218.  It is known that the signature of shock break-out is a hot black-body X-ray spectrum immediately after the explosion, and C06 argued that the hot temperature of the thermal X-ray component is indicative of radiation emitted by a shock-heated plasma.  Furthermore, the characteristic radius of the emitting region was $\sim 5 \times 10^{12}$ cm (Waxman el al. 2007 also find the emitting region to be at a radius of $7.8 \times 10^{12}$ cm), which though is quite large, (i.e. of order of the radius of a blue supergiant), was explained by C06 and Waxman et al. (2007) by the presence of a massive, dense stellar wind surrounding the progenitor, which is common for Wolf-Rayet stars (the supposed progenitors of type Ibc SNe and possibly long-duration GRBs).  Both authors argue that the thermal radiation is observed once the shock, that is driven into the wind, reaches a radius where the wind becomes optically thin.

For XRF 100316D / SN~2010bh, the temperature of the black-body component in the X-ray spectrum implies an emitting radius of $\sim 8 \times 10^{11}$ cm, which is almost an order of magnitude larger than the radius of a Wolf-Rayet star ($\sim 10^{11}$ cm).  As for XRF 060218, the large radius of the X-ray emitting region could be the result of the shock being driven into the dense stellar wind, and the thermal radiation is observed once the wind density decreases and becomes optically thin.

Thus, the brightness of the $B$-band detection at $\rm{t-t_{o}}=0.598$ days, when taken in tandem with the hard value of the power-law index ($\beta =+0.94 \pm 0.05$) of the SED at this epoch, as well as the presence of the extremely hot thermal component ($T \approx 1.6 \times 10^{6}$ K) in the X-ray spectrum at $144$ s $ \rm \le t-t_{o} \le 737$ s, suggests that we have detected optical emission coming from the cooling, expanding envelope that was heated by the passage of the shock-wave through it.

\subsection{Stretch Factor relative to SN~1998bw}

We have determined the stretch factor ($s$) and luminosity factor ($k$) of SN~2010bh in relation to the archetype GRB-SN, SN~1998bw.  To do this we have created synthetic flux SN~1998bw light curves in filters $BVR_{c}i$ as they would appear if they occurred at $z=0.0591$.  At each epoch, an SED of the observed SN~1998bw light curves is built and interpolated using polynomials.  This allows us to cover the temporal-frequency space for SN~1998bw and build synthetic light cuvres for other specified observed frequencies.  When constructing the synthetic LCs we have accounted for the inverse-squared distance (i.e. luminosity distance) suppression of flux, as well as rest-frame extinction and time dilation effects.  We then fit the flux SN~1998bw light curves with an empirical relation 
of the form:

\begin{center}
\begin{equation}
\label{equ:stretch}
U(t) = A + st\left(\frac{e^{(\frac{-t^{\alpha_{1}}} F)}}{1 + e^{(\frac{p-t} R) }}\right) + t^{\alpha_{2}}\log(t^{-\alpha_{3}})
\end{equation}
\end{center}

\noindent where $A$ is the intercept of the line and $s$ and $F$ are related to the respective amplitude and width of the function.  The exponential cut-off function for the rise has a characteristic time $R$ and a phase zero-point $p$, while $\alpha_{1}$, $\alpha_{2}$ and $\alpha_{3}$ are free parameters.  All of the parameters are allowed to vary during the fit.

Once the flux light curve of SN~1998bw was fit by Equation (\ref{equ:stretch}), the stretch and luminosity factors of SN~2010bh were determined by fitting the following equation:  

\begin{center}
\begin{equation}
\label{equ:stretch2}
 W(t)=k \times U(t/s)
\end{equation} 
\end{center}

\noindent to the flux light curve of SN~2010bh in each filter. 

The stretch factors of SN~2010bh relative to SN~1998bw were determined using data taken several days after the initial burst ($\rm{t-t_{o}} \ge 7.0$ days) as the early light curves, whose shapes are altered by light coming from an additional shock break-out component, are very different to that of SN~1998bw.  The results of the fit are listed in Table \ref{table:2010bh_stretch}, where the quoted errors are statistical only.

\begin{table}
\begin{center}
\begin{minipage}{80mm}
\centering
\caption{Luminosity ($k$) and Stretch ($s$) Factors of SN~2010bh \label{table:2010bh_stretch}}
\begin{tabular}{ccc}
\hline
 Filter & $s$ & $k$ \\
\hline
$B$ & $0.62\pm0.01$ & $0.41\pm0.01$ \\
$V$ & $0.64\pm0.01$ & $0.43\pm0.01$ \\
$R_{c}$ & $0.62\pm0.02$ & $0.40\pm0.01$ \\
$i$ & $0.60\pm0.01$ & $0.48\pm0.01$ \\
\hline
\end{tabular}

\medskip
\end{minipage}
\end{center}
\end{table}

For GRB and XRF events where an optically bright SN has been detected (i.e. not including events such as SN-less GRBs 060505 and 060614; e.g. Fynbo et al. 2006), it is seen that SN~2010bh is the faintest SN to date that has been linked spectroscopically with a GRB or XRF (see Table \ref{table:GRB_stretch}), and is of similar peak brightness as the SN associated with GRB 970228 ($k=0.40\pm0.29$, though the host extinction is unknown), but not as faint as the (possible) supernova that was photometrically-linked to GRB 101225A ($k=0.08\pm0.03$, Th\"one et al. 2011).  The slower evolution of SN~2006aj relative to SN~2010bh seen upon inspection of the optical light curves are echoed in the computed stretch factors for the two SNe.  In the $B$ filters they are approximately the same value (SN~2010bh: $s=0.62 \pm 0.01$; SN~2006aj: $s=0.60 \pm 0.01$), however in the redder filters SN~2010bh has smaller stretch factors relative to SN~2006aj.

It should be mentioned that SN~1998bw is not the ideal template for SN~2010bh.  At early times light from the shock break-out changes the shape of the light curve, which affects how well SN~1998bw, for which no shock break-out was observed, can be used as a template.  Even at late times it is seen in Figure \ref{fig:GRB100316D_LCs} that SN~2010bh decays more slowly than the stretched LC of SN~1998bw, thus showing that the two SNe evolve quite differently.  The limitations of using SN~1998bw as template was also noted by Ferrero et al. (2006), where they used an additional power-law component when fitting their SN~1998bw template to their multi-filter observations of SN~2006aj.

\section{Reddening \& Absolute Magnitudes}
\label{sec:ebv}

Determining the absolute magnitude of SN~2010bh requires an estimate of the amount of extinction local to the site of the SN explosion.  
Drout et al. (2010)
found for a sample of local type Ibc SNe that the mean values of the $(V-R)$ colour at ten days after maximum are tightly distributed.  The authors found at ten days after $V$ maximum $\langle(V-R)_{V10}\rangle=0.26 \pm 0.06$ mag, and ten days after $R$  maximum $\langle(V-R)_{R10}\rangle=0.29 \pm 0.08$ mag.  For SN~2010bh we found the $(V-R)$ colour at ten days after $V$ and $R$ maximum to be, respectively, $\langle(V-R)_{V10}\rangle=0.38 \pm 0.05$ mag, and $\langle(V-R)_{R10}\rangle=0.43 \pm 0.05$ mag.  The difference in the colours of the sample found by Drout et al. (2010) in comparison with SN~201bh implies a colour excess of $E(V-R) \simeq 0.12$ mag, which in turn implies $E(B-V) \simeq 0.18$ mag (using the relative extinction values in Table 6 of Schlegel et al. 1998).

For comparison, using the Balmer H$\alpha$ and H$\beta$ lines fluxes to measure the Balmer decrement, 
Starling et al (2011) found a combined reddening (host and foreground) of $E(B-V)=0.178$ mag for a nearby HII region (``Source A''; see Figure 11 in Starling et al. 2011).  The foreground extinction found from the dust maps of Schlegel et al (1998) is $E(B-V)= 0.117$ mag, which implies a rest-frame, host extinction of $E(B-V) = 0.061$ mag.  This value is consistent with the value found via comparing the colour curves, albeit somewhat smaller, but provides additional credence to the method described by Drout et al. (2010).  We note that the value for the colour excess found by Starling et al. (2011) was not for the specific explosion site of SN~2010bh as the SN was still bright during their observations, thus the authors made the colour excess measurement of a nearby bright knot/blob instead.  None the less, these independent analysis imply a small colour excess for the location near and around SN~2010bh.  Using the value $E(B-V) = 0.18 \pm 0.08$ mag, we have calculated the absolute magnitudes for SN~2010bh (Table \ref{table:2010bh_phot_params}).

In comparison with the peak magnitudes of almost all of the previously detected GRB-SN tabulated by Cano et al. (2011), it is seen that the peak absolute $V$-band magnitude of SN~2010bh ($M_{V} = -18.62 \pm 0.08$) is the faintest ever detected for a spectroscopically-detected GRB-SNe and is $\sim 0.8$ mag fainter than SN~1998bw at peak brightness.  Cano et al. (2011) found for a sample of 22 GRB-SNe (the authors excluded events where only an upper limit to the SN brightness was obtained; e.g. GRBs 060505 \& 060614, Fynbo et al. 2006), where measurements of the host extinction have been made and included in the absolute magnitude calculation, an average, absolute, $V$-band magnitude of $\langle M_{V}\rangle = -19.00$, with a standard deviation of $\sigma = 0.77$ mag.  In a separate analysis of the peak GRB-SNe magnitudes, Richardson (2009) determined for a sample of 14 GRB-SNe with known values for the host extinction, an average absolute $V$-band magnitude of $\langle M_{V}\rangle = -19.20$, with a standard deviation of $\sigma = 0.70$ mag.  In comparison with these analyses it appears that SN~2010bh lies at the lower end of the peak brightness distribution.

\section{Explosion parameters}
\label{sec:bol}

\begin{figure}
\centering
\includegraphics[scale=0.8, angle=0]{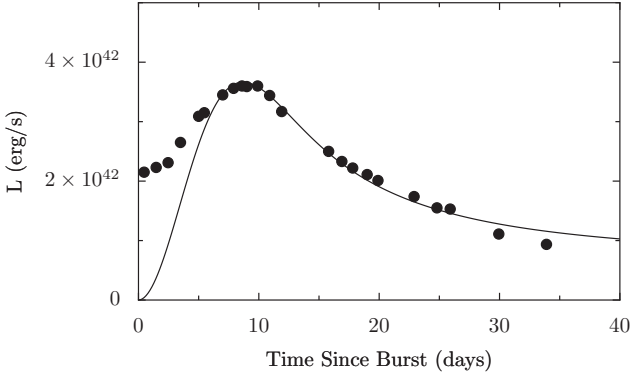}
\caption{The quasi-bolometric light curve for XRF 100316D / SN~2010bh in the $\rm 3,000 \AA - 16,600 \AA$ wavelength range, which was obtained by integrating the flux in the optical and infrared filters $UBVRiJH$.  The model (solid line) that is used to determine the nickel mass and ejecta mass is also plotted.  Peak bolometric light is found to be at $t-t_{\rm o} = 8.57 \pm 0.04$ days.}
\label{fig:GRB100316D_Bol}
\end{figure}

We have constructed a quasi-bolometric light curve of XRF 100316D / SN~2010bh (Figure \ref{fig:GRB100316D_Bol}) in the $\rm 3,000 \AA - 16,600 \AA$ wavelength range (i.e. $UBVR_{c}iJH$) by taking the following steps:  (1) converting all magnitudes to monochromatic fluxes after correcting for foreground and host extinction; (2) the SED at each epoch was integrated over frequency.  At epochs where there was no contemporaneous data, the individual light curves in each filter were extrapolated after fitting Equation (\ref{equ:stretch}) to the individual light curves in each filter.  For epochs where there was no IR data, we estimated the amount of IR flux as a fraction of the total bolometric flux using the existing data and extrapolating between epochs. 

For epochs before the first IR detection (i.e. $\rm{t-t_{o}}=8.6$ days), we have assumed a constant fraction of flux at IR wavelengths of $20\%$.  We note that this assumption of a constant amount of flux at IR wavelengths is valid only for the first dozen or so days, as it is seen that the fraction of light at IR wavelengths actually increases with time and can be as high as $45\%$ at late times (see Figure \ref{fig:GRB100316D_Bol_percentage} in this work, as well as Figure 9 of Modjaz et al. 2009).  The increase in the amount of flux at redder wavelengths is expected however, as it is seen that core-collapse SNe become redder over time.  We estimate the uncertainty in the bolometric magnitude, which is dominated by systematic errors described in the method above, to be of order $0.2$ mag.

We have calculated the time of peak bolometric light by fitting a series of different order polynomials to the bolometric light curve.  We find peak bolometric light to be at $\rm{t-t_{o}} = 8.57 \pm 0.04$ days, where the error is statistical and determined from the distribution of peak times determined from the different order polynomials.

Here we have determined a quasi-bolometric light curve of XRF 100316D / SN~2010bh up to $\approx 35$ days.  This time interval corresponds to the \emph{photospheric phase}.  To attempt to extract physical parameters of the progenitor star we have used a simple analytical model that was developed by Arnett (1982) for type I SNe, which was later expanded by Valenti et al. (2008).  

Arnett-like models have been used by many authors to intrepret the light curves of numerous Ibc SNe and GRB-SNe events, including: Ib SN~1999dn, Benetti et al. (2011); Ic-BL SN~2003jd, Valenti et al. (2008); Ic SN~2004aw, Taubenberger et al. (2006); Ic-BL SN~2009bb, Pignata et al. (2011); GRB-SNe 1998bw, Iwamoto et al. (1998); Richardson et al. (2006) for a sample of Ibc SNe; Drout et al. (2011) for a sample of Ibc SNe.  The model assumes: (1) homologous expansion of the ejecta, (2) spherical symmetry, (3) all of the radioactive nickel ($^{56}$Ni) is located at the centre of the explosion and does not mix, (4) radiation-pressure dominance, (5) a small initial radius before explosion ($\rm{R_{o}} \rightarrow 0$), and (6) the applicability of the diffusion approximation for photons (i.e. the \emph{photospheric phase}).  We also assume a constant opacity $\kappa=0.07$ cm$^{2}$g$^{-1}$ (e.g. Chugai 2000), which is justified if electron scattering is the dominant opacity source\footnote{Metal lines also contribute to the total opacity, and strictly speaking the electron scattering opacity is not actually constant in time as the ionization state changes.  For triply ionized ejecta, $\kappa \approx 0.07$.} (e.g. Chevalier 1992).  Finally, with respect to Arnett (1982) who only considered the energy produced by the decay of nickel into cobalt ($^{56}$Ni $\rightarrow$ $^{56}$Co), we also include, using the methodology described in Valenti et al. (2008), the energy produced by the decay of cobalt into iron ($^{56}$Co $\rightarrow$ $^{56}$Fe):\\ \ \\

\noindent $\rm{L(t) = M_{Ni}e^{-x^2}} \times $

\begin{center}
\begin{equation}
\label{equ:bol1}
\rm{\left((\epsilon_{Ni} - \epsilon_{Co}) \int_{0}^{x}A(z)dz+ \epsilon_{Co}\int_{0}^{x}B(z)dz\right)}
\end{equation} 
\end{center}
%
where 
\begin{center}
\begin{equation}
\label{equ:bol2}
A(z)=2ze^{-2zy+z^2}, B(z)=2ze^{-2zy+2zs+z^2} 
\end{equation} 
\end{center}
%
and $x\equiv t/\tau_{m}$, $y\equiv \tau_{m}/(2\tau_{Ni})$, and $s\equiv (\tau_{m}(\tau_{Co}-\tau_{Ni})/(2\tau_{Co}\tau_{Ni}))$.  The energy release in one second by one gram of $^{56}$Ni and $^{56}$Co are, respectively, $\epsilon_{Ni}=3.90 \times 10^{10}$ erg s$^{-1}$ g$^{-1}$ and $\epsilon_{Co}=6.78 \times 10^{9}$ erg s$^{-1}$ g$^{-1}$ (Sutherland \& Wheeler 1984; Cappellaro et al. 1997).  The decay times of $^{56}$Ni and $^{56}$Co, respectively, are $\tau_{Ni}=8.77$ days (Taubenbeger et al. 2006 and references therein) and  $\tau_{Co}=111.3$ days (Martin 1987).

$\tau_{m}$ is the effective diffusion time and determines the width of the bolometric light curve.  $\tau_{m}$ is expressed in relation to the  opacity $\kappa$ and the ejecta mass $\rm{M_{ej}}$, as well as the photospheric velocity $\rm{v_{ph}}$ at the time of bolometric maximum and the kinetic energy of the ejecta $\rm{E_{k}}$:

\begin{center}
\begin{equation}
\label{equ:tau}
\tau_{m} \approx \left(\frac{\kappa}{\beta c}\right)^{1/2} \left(\frac{\mathrm{M^{3}_{ej}}}{\mathrm{E_{k}}}\right)^{1/4}
\end{equation} 
\end{center}
%
where $\beta \approx 13.8$ is a constant of integration (Arnett 1982).  Thus, for SNe with similar ejecta velocities, a wider bolometric light curve implies more ejecta mass and higher kinetic energies, while for SNe with similar amount of ejected material, a wider light curve implies slower ejecta velocities.

We have fit the data at times $> 5$ days (as to not include flux that may also be coming from the shock-heated, expanding stellar envelope).  We have taken the photospheric velocity around peak light for XRF 100316D / SN~2010bh to be $\rm{v_{ph}} \approx 25,000$ km s$^{-1}$ (determined from the Si II absorption lines measured by Chornock et al. 2010; their Figure 3), the analytical model yields $\rm{M_{Ni}}=0.10 \pm 0.01 M_{\odot}$, $\rm{M_{ej}}=2.24 \pm 0.08 M_{\odot}$, $\rm{E_{k}} = 1.39 \pm 0.06  \times 10^{52}$ erg.  Our results are displayed in Table \ref{table:Ic_bol_params}, where they are compared with other Ibc and GRB-SNe.


The quoted errors are statistical and include the uncertainties in the
photometry, extinction, and calibration.  However, the dominant sources of
error are systematic.  For example, our assumption of a
constant value of the opacity, is perhaps valid during the photospheric
phase, but may differ in value from our assumed value of $\kappa=0.07$
cm$^{2}$g$^{-1}$.  Using Equation \ref{equ:tau} and the values of the explosion energy
and ejecta mass from the literature for SN~1998bw, we calculate an opacity of $\kappa \approx 0.05$ cm$^{2}$g$^{-1}$.  Since
the total opacity depends on the metal content of the ejecta as well as that from electron scattering, it is quite possible that the value of the opacity may have a slightly higher or lower value than what we have used here.   A small change in the value of the opacity leads to an uncertainity in the ejected mass of ($\rm{M_{ej}}$) of $25 \% -30 \%$ and an uncertainity in the kinetic energy ($\rm{E_{k}}$) of $60 \%-75\%$.  The uncertainty in the nickel mass is less, as it is mainly affected by the uncertainty in the rest-frame extinction, and to a lesser extent in the photometric errors.

As an alternative method to check how much mass was ejected during SN~2010bh, we have compared the width of bolometric LC of XRF 100316D / SN~2010bh with those of three other spectroscopically-confirmed GRB-SNe: SN~1998bw, SN~2003dh and SN~2006aj.  If we assume that the opacity is the same in each event, we can solve Equation (\ref{equ:tau}) to find the ejected mass for SN~2010bh:

\begin{center}
\begin{equation}
\label{equ:bolmass}
\rm{M_{ej,10bh} \approx M_{ej,GRB} \left(\frac{v_{10bh}}{v_{GRB}}\right) \left(\frac{\tau_{m,10bh}}{\tau_{m,GRB}}\right)^{2}} 
\end{equation} 
\end{center}

First, using SN~1998bw, from the literature (e.g. Iwamoto et al. 1998; Patat et al. 2001) it is expected that $\approx 8-10 M_{\odot}$ of material was ejected.  We have measured $\tau_{m,98bw}$ from our bolometric LC of SN~1998bw (see Section 7.2, where we have also measured $\tau_{m}$ for SN~2006aj and SN~2003dh on way to deriving physical parameters using our analytical model) and thus derive $\rm{M_{ej,10bh}} \approx 3.4 M_{\odot}$.  For SN~2003dh, it is expected that $\approx 7 \pm 3 M_{\odot}$ of material was ejected (e.g. Deng et al. 2005), which in turn implies for SN~2010bh an ejected mass of $\rm{M_{ej,10bh}} \approx 3.3 M_{\odot}$.  For SN~2006aj, it was determined that $\approx 2 M_{\odot}$ of material was ejected during the explosion (e.g. Mazzali et al. 2006).  This implies an ejected mass for SN~2010bh of $\rm{M_{ej,10bh}} \approx 2.7 M_{\odot}$.  In turn, these values for the ejected mass imply, respectively, explosion energies for SN~2010bh of $\rm{E_{k,10bh}}\approx 2.1 \times 10^{52}$ erg, $\rm{E_{k,10bh}}\approx 2.0 \times 10^{52}$ erg, and $\rm{E_{k,10bh}}\approx 1.7 \times 10^{52}$ erg.  This method gives larger values for the ejected mass of SN~2010bh which imply larger explosion energies.  On average, these values are larger than those determined directly from our fit. 
%
%

It is important to consider the limitations of our model and the implied caveats from our assumptions.  While it appears that our model is able to accurately reproduce the nickel masses in each of these events, the energy estimates have probably the largest uncertainities.  As mentioned, Arnett (1982) considers spherical geometry, while GRB-SNe are thought to be asymmetric and may well have an unusual distribution of nickel mass in the ejecta mass.  Furthermore, this model calculates an explosion energy by assuming that all of the ejecta are moving at the photospheric velocity.  In the case of XRF 100316D / SN~2010bh, while Chornock et al. (2010) observed absorptions in the spectra at $25,000$ km s$^{-1}$, it does not mean that \emph{all} of the ejecta are moving at that velocity.  It is quite possible that only a small or modest fraction of the mass is expelled at high velocities (either isotropically or collimated into a jet) creating the absorption lines in the spectra, while most of the mass moves at slower velocities.  A more accurate way of determining the actual explosion energy would be to understand the way the mass is distributed in the ejecta (i.e. via detailed modeling), and integrate over the density-velocity profile.

\section{Discussion and Further Analysis}
\label{sec:discuss}

\subsection{SN~2010bh in relation to SN~2006aj}
\label{sec:2010bh_2006aj}

\begin{figure}
\centering
\includegraphics[scale=0.35, angle=-90]{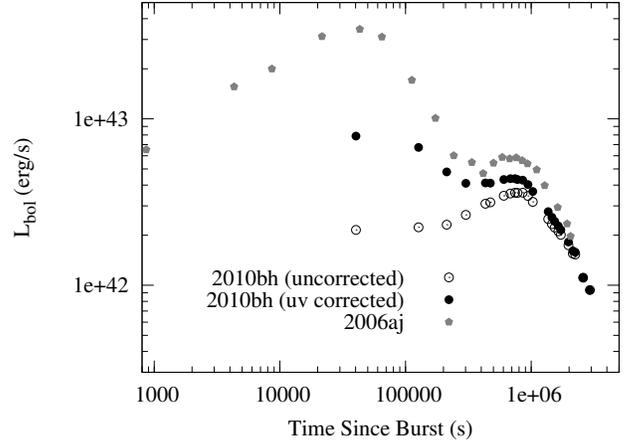}
\caption{The UVOIR bolometric light curves, in the $\rm 1,900\AA -16,660 \AA$ wavelength range, of XRF 100316D / SN~2010bh (filled circles) and XRF 060218 / SN~2006aj (grey hexagons). When constructing the bolometric light curve of SN~2010bh, we have assumed that the same fraction of the bolometric flux emitted at UV wavelengths is the same for SN~2010bh and for SN~2006aj.  Plotted for comparison is the $UBVR_{c}iJH$ bolometric light curve of SN~2010bh (open circles).  All photometry has been corrected for foreground and host extinction.}
\label{fig:GRB100316D_Bol_2006aj}
\end{figure}

\begin{figure}
\centering
\includegraphics[scale=0.35, angle=-90]{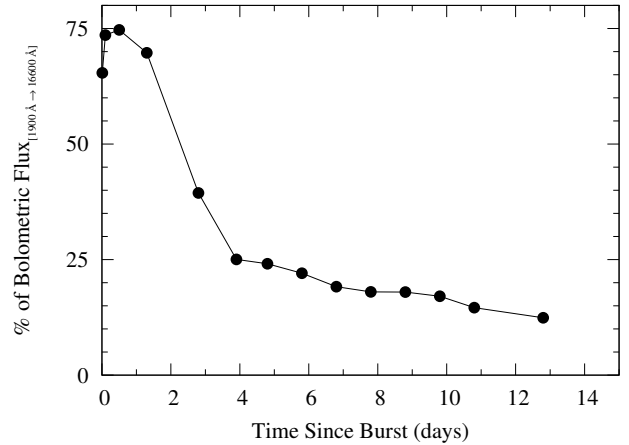}
\caption{XRF 060218 / SN~2006aj: The amount of UV flux as a fraction of the total bolometric flux in the $\rm 1,900\AA -16,660 \AA$ wavelength regime, plotted over time.  It is seen that up to a few days after the initial explosion, the amount of UV flux is considerable.  More importantly, even at late times ($\ge 10$ days) the amount of flux at UV wavelengths is a non-negligible fraction of the total bolometric flux.}
\label{fig:GRB100316D_Bol_2006aj_UV}
\end{figure}

\begin{figure}
\centering
\includegraphics[scale=0.35, angle=-90]{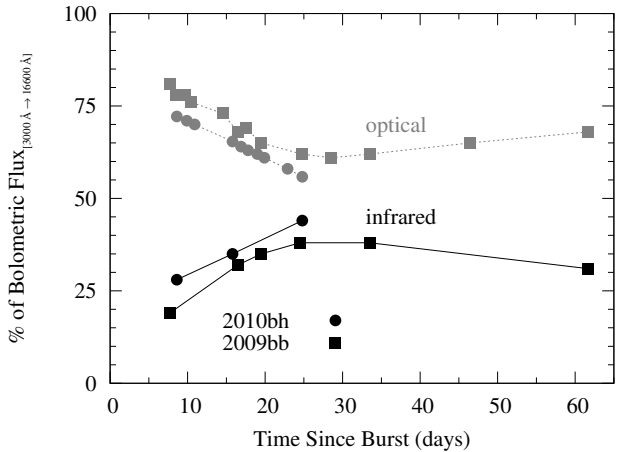}
\caption{The amount of flux as a fraction of the total bolometric flux (in filters $UBVR_{c}iJH$) that is obtained by integrating the optical (grey, dashed-line) and infrared (black, solid-line) flux for SN~2010bh and SN~2009bb.  It is seen that the behaviour is similar in each SN event.}
\label{fig:GRB100316D_Bol_percentage}
\end{figure}

It was shown in the preceeding section that, despite ejecting only a moderate amount of material, SN~2010bh was an energetic explosion, mainly due to the large photospheric velocities.  To put the results of our bolometric light curve modelling into context, we first compare SN~2010bh with SN~2006aj and then with the other GRB-SNe.

The ultra-violet, optical and infrared (UVOIR) bolometric light curve (in the $\rm 1,900\AA-16,660 \AA$ wavelength regime) of XRF 060218 / SN~2006aj, which has been constructed from the data published in the literature (Campana et al. 2006; Sollerman et al. 2006; Cobb et al. 2006; Ferrero et al. 2006), is displayed in Figure \ref{fig:GRB100316D_Bol_2006aj}.  It is seen that during the first few days the fractional contribution of UV flux (i.e. light detected in the $\rm 1,900\AA-3,000 \AA$ wavelength regime) to the total bolometric LC (Figure \ref{fig:GRB100316D_Bol_2006aj_UV}) is considerable due to the shock break-out of the progenitor star (Campana et al. 2006), which decays at a linear rate after $\approx 4$ days.  It is important to note that the contribution of flux at UV wavelengths is \textbf{not negligible}, and at roughly $10$ days still contributes 15\% of the total bolometric flux.

By making the assumption that the fraction of UV flux of the total bolometric flux at a given epoch is the same in SN~2010bh as it is in SN~2006aj, we have used SN~2006aj as a template/proxy for creating a ``UV-corrected'' bolometric light curve for SN~2010bh, which is also displayed in Figure \ref{fig:GRB100316D_Bol_2006aj}.  The bolometric light curve of SN~2010bh is shown twice: (1) first, by considering flux only from the optical ($UBVR_{c}i$) and infrared ($JH$) filters and, (2) by including a contribution of flux at UV wavelengths.

Our assumption that the fraction of UV flux is similar in each event is perhaps not unreasonable when we also consider the percentage of the bolometric flux detected in the optical and IR regimes for SN~2010bh and SN~2009bb.  Plotted in Figure \ref{fig:GRB100316D_Bol_percentage} is a plot of the amount of light detected in the optical and infrared as a fraction of the total bolometric flux in each epoch for SN~2009bb and SN~2010bh.  The data included in this plot does not include all of the data obtained for these two events, but only those where there is contemporaneous optical and IR data.  The first IR detection of SN~2010bh was during the first \emph{HST} epoch at $\rm{t-t_{o}}=8.6$ days, and the first IR detection of SN~2009bb was at $\rm{t-t_{o}}=7.7$ days (Pignata et al. 2011).   Inspection of the optical and infrared filters pairs reveal similar behaviour in both events, though it is seen that more flux is emitted at IR wavelengths in SN~2010bh relative to SN~2009bb.  

The fractional contribution of the optical and IR flux to the bolometric light curves for these two events, as well as the large UV contribution seen in SN~2006aj, is similar to that seen in XRF 080109 / SN~2008D.  Modjaz et al. (2009) also compared the relative contribution of the UV, optical, and IR flux in the bolometric LC of SN~2008D for the first $\approx 30$ days (Figure 9; Modjaz et al. 2009), and noted a large contribution of flux from 2008D at UV wavelengths ($\approx 40\%$ at $\rm{t-t_{o}}\approx 0.8$ days).  This is not quite as high as for SN~2006aj which is $\approx 70\%$ at $\rm{t-t_{o}}\approx 0.8$ days, but their UV contribution does not include far-UV flux, whereas SN~2006aj does.  Indeed at $\rm{t-t_{o}}=1.8$ days, when they do include the far-UV flux, the total UV flux is seen to increase.  Modjaz et al. (2009) also showed that the amount of IR flux increased over time, while the optical flux increased linearly for the first ten days or so, then decreased.  While there is scant data plotted in Figure \ref{fig:GRB100316D_Bol_percentage} before $\rm{t-t_{o}} = 10$ days, our plot does show that the optical contribution decreases during $8.6$ days $ \le \rm{t-t_{o}} \le 24.8$ days for SN~2010bh, whereas for SN~2009bb, decreases until $\approx 25-30$ days, then appears to contribute a nearly constant, if slightly increasing, amount to the total bolometric flux.  Indeed, when comparing Figure \ref{fig:GRB100316D_Bol_percentage} with Figure 9 of Modjaz et al. (2009), the general behaviour of SNe 2008D, 2009bb and 2010bh appear to follow similar trends in their relative bolometric fluxes during the first $\approx 30$ days.

Next, we applied our analytical model to our constructed bolometric light curve of XRF 060218 / SN~2006aj.  Taking the peak photospheric velocity to be $\rm{v_{ph}}\approx 20,000$ km s$^{-1}$ (Pian et al. 2006), we find: $\rm{M_{Ni}}=0.16 \pm 0.01 M_{\odot}$, $\rm{M_{ej}}=1.63 \pm 0.05 M_{\odot}$, $\rm{E_{k}} = 6.47 \pm 0.20 \times 10^{51}$ erg.  In comparison, Mazzali et al. (2006) found when modeling the spectra and photometry of SN~2006aj parameters of $\rm{M_{Ni}} \approx 0.2 M_{\odot}$, $\rm{M_{ej}} \approx 2 M_{\odot}$, $\rm{E_{k}} \approx 2 \times 10^{51}$ erg.  It appears that our simple model is consistent with the values of the physical parameters of the SN~2006aj explosion determined via detailed modeling, but over-estimates the explosion energy by a factor of $\sim 3$.  As mentioned in Section (5), our estimate of the energy assumes that all of the mass moves at the photospheric velocity.  Here is a clear example where this assumption \emph{over}-estimates the explosion energy, implying that a bulk of the ejecta are moving at slower velocities. 

Alternatively, if we exclude the contribution of flux at UV wavelengths, similar to the $UBVR_{c}I_{c}JH$ bolometric LCs presented by Pian et al. (2006), we find for SN~2006aj: $\rm{M_{Ni}}=0.14 \pm 0.01 M_{\odot}$, $\rm{M_{ej}}=1.92 \pm 0.12 M_{\odot}$, $\rm{E_{k}} = 7.64 \pm 0.49 \times 10^{51}$ erg, which is again consistent with those found by Mazzali et al. (2006), but the explosion energy is again over-estimated, this time by a factor of $\sim 4$.  In both cases the errors are estimated as per Section (5).  We note that by including the contribution of flux at UV wavelengths, which is considerable at early times (i.e. $\le 5$ days), this increases the total luminosity of the SN, but makes the bolometric light curve peak earlier, thus decreasing the overall width of the LC (i.e. less mass is ejected).  The results of our modelling are displayed in Table \ref{table:Ic_bol_params}.

When we apply our model to the bolometric light curve of 2010bh that includes a UV correction, we find $\rm{M_{Ni}}=0.12 \pm 0.01 M_{\odot}$, $\rm{M_{ej}}=1.93 \pm 0.07 M_{\odot}$, $\rm{E_{k}} = 1.20 \pm 0.05 \times 10^{52}$ erg.  The inclusion of the UV flux makes the overall bolometric light curve brighter, implying more $^{56}$Ni was created during the explosion.  However, as for SN~2006aj, the inclusion of the UV flux, which is dominant at early times, causes the light curve to peak earlier, thus making the overall shape of the bolometric light curve narrower, and implying that less mass was ejected during the explosion (and thus a less energetic explosion).

Never the less, whether an estimated contribution of the UV flux is included or not, the results of bolometric LC modeling imply that SN~2010bh ejected a mass of $\rm{M_{ej}}=1.9-2.2 M_{\odot}$, has a nickel mass of $\rm{M_{Ni}}=0.10 - 0.12 M_{\odot}$, and a kinetic energy of roughly $\rm{E_{k}} = 1.2-1.4 \times 10^{52}$ erg.

\begin{table*}
\begin{center}
\caption{Physical parameters of Ibc SNe and GRB-SNe \label{table:Ic_bol_params}}
\begin{tabular}{ccccccccc}
\hline
 SN & type & $\rm{v_{ph}}$ (km s$^{-1}$) & $\rm{M_{Ni}}$ ($M_{\odot}$)  & $\rm{M_{ej}}$ ($M_{\odot}$) & $\rm{E_{k}}$ ($10^{52}$ erg) & Wavelength Range ($\rm \AA$) & Ref. & Note$^{a}$ \\
\hline
1994I & Ic & 10,000 & $0.07\pm0.01$  & $0.88\pm0.05$ & $0.09\pm0.01$ & $4,400 - 8,000$ & (1) & \textbf{This Work}\\
1994I & Ic & 10,000 & $\approx 0.07$ & $\approx0.9$  & $0.10\pm0.01$ & $4,400 - 8,000$ & (12) & Literature \\
GRB 980425 / 1998bw & Ic-BL & 18,000 & $0.43\pm0.02$  & $7.04\pm0.57$  & $2.27\pm0.17$ & $3,000 - 16,600$ & (2), (3), (4), (5) & \textbf{This Work}  \\
GRB 980425 / 1998bw & Ic-BL & 18,000 & $0.4-0.7$ & $8 \pm 2$ & $2-5$  & $3,000 - 16,600$ & (13), (14) (5) & Literature \\
GRB 980425 / 1998bw & Ic-BL & 18,000 & $0.52\pm0.02$  & $5.36\pm0.53$  & $1.73\pm0.17$ & $1,900 - 16,600$ & (2), (3), (4), (5)& \textbf{This Work} \\
1999dn & Ib & - & $\approx 0.11$ & $4-6$ & $0.5$ & $3,000 - 16,600$& (18) & Literature\\
1999ex & Ib & - & $0.25$ & $0.9$ & $0.03$ & - & (16) & Literature\\
2002ap & Ic-BL & - & $\approx 0.07$ & $2.5-5.0$ & $0.4-1.0$ & $3,000 - 16,600$ & (17) & Literature \\
GRB 030329 / 2003dh & Ic-BL &  20,000 & $0.39\pm0.03$  & $4.82\pm0.54$  & $1.92\pm0.21$ & $3,000 - 8,000$  & (6)& \textbf{This Work} \\
GRB 030329 / 2003dh & Ic-BL &  20,000 & $\approx 0.4$  & $7\pm3$  & $3.5\pm1.5$ & $3,000 - 8,000$  & (6) & Literature \\
2003jd & Ic-BL & - & $0.36\pm0.04$ & $3.0\pm0.5$ & $\approx 0.7$ & $4,400 - 8,000$ & (19) & Literature \\
2004aw & Ic & - & $0.2-0.3$ & $3-8$ & $\approx 0.8$ &  $3,000 - 16,600$ & (19), (20) & Literature \\
XRF 060218 / 2006aj & Ic-BL &  20,000 & $0.14\pm0.01$  & $1.92\pm0.12$  & $0.76\pm0.05$ & $3,000 - 16,600$  & (7), (8), (9), (10)& \textbf{This Work} \\
XRF 060218 / 2006aj & Ic-BL &  20,000 & $\approx 0.2$  & $\approx 2$  & $\approx 0.2$ & $3,000 - 16,600$  & (15) & Literature \\
XRF 060218 / 2006aj & Ic-BL &  20,000 & $0.16\pm0.01$  & $1.63\pm0.05$  & $0.65\pm0.02$ & $1,900 - 16,600$  & (7), (8), (9), (10)& \textbf{This Work} \\
2009bb & Ic-BL &  15,000  & $0.23\pm0.01$  & $3.86\pm0.13$ & $1.73\pm0.06$  & $3,000 - 16,600$  & (11) & \textbf{This Work} \\ 
2009bb & Ic-BL &  15,000  & $0.22\pm0.06$  & $4.1\pm1.9$ & $1.8\pm0.7$  & $3,000 - 16,600$  & (11) & Literature \\ 
2009jf & Ib & - & $0.17\pm0.03$ & $4-9$ & $0.3-0.8$ & $3,000-8,000$ & (21) & Literature \\
XRF 100316D / 2010bh & Ic-BL &  25,000 & $0.10\pm0.01$ & $2.24 \pm 0.08$  & $1.39 \pm 0.06$ & $3,000 - 16,600$ & \textbf{This Work} & - \\
XRF 100316D / 2010bh & Ic-BL &  25,000 & $0.12\pm0.01$ & $1.93 \pm 0.07$  & $1.20 \pm0.05$ & $1,900 - 16,600$ & \textbf{This Work} & - \\
\hline
\end{tabular}
\medskip\\

$^{a}$ Physical parameters determined in: (1) this paper, or (2) taken from the literature.\\
(1) \cite{Richmond96}, (2) \cite{Galama98}, (3) \cite{McKenzie1999}, (4) \cite{Sollerman02}, (5) \cite{Patat01}, (6) \cite{Deng05}, (7) \cite{Campana2006}, (8) \cite{Sollerman06}, (9) \cite{Cobb06}, (10) \cite{Ferrero06}, (11) \cite{Pignata2011}, (12) \cite{Iwamoto94}, (13) \cite{Iwamoto98}, (14) \cite{Nakamura01}, (15) \cite{Mazzali2006}, (16) \cite{Richardson2006} and references therein, (17) \cite{Mazzali2002}, (18) \cite{Benetti2011}, (19) \cite{Valenti08}, (20) \cite{Taubenberger06}, (21) \cite{Sahu2011}
\end{center}
\end{table*}

\subsection{Bolometric Light Curves of other GRB-SNe and local stripped-envelope SNe}

It is seen that, despite the similaries of the high-energy properties of XRF 100316D and XRF 060218 (Starling et al. 2011), the accompaning supernova are somewhat different, with SN~2010bh being much more energetic.  In this section we compare the bolometric properties of SN~2010bh with other GRB-SNe, as well as non-GRB-SNe: Ic SN~1994I and broad-lined Ic SN~2009bb.  We investigate whether clear differences exist in the bolometric properties of stripped-envelope, core-collapse SNe that produce a GRB and those that do not.

To date, numerous authors have determined physical parameters of Ibc SNe by modeling photometry \emph{and} spectroscopy.  Many authors employ analytical models that are derived from the original prescription of Arnett (1982) to describe bolometric LCs, however very often these models only supplement the derivation of physical parameters that are determined by comparing synthetic spectra created by synthesis codes to observations.  In this work we are attempting to obtain physical parameters of previously-studied events by modeling only the bolometric LCs, and compare our results with those obtained via more detailed analyses.

\begin{figure}
\centering
\includegraphics[scale=0.35, angle=-90]{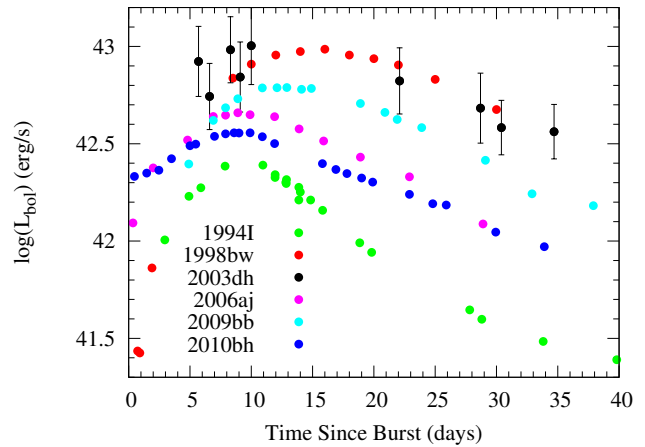}
\caption{The quasi-bolometric $UBVR_{c}iJH$ light curves of XRF 100316D / SN~2010bh as well as three spectroscopically-confirmed GRB-SNe: 1998bw, 2003dh \& 2006aj and two local type Ic SNe that were not accompanied by a GRB-trigger: Ic SN~1994I \& broad-lined Ic SN~2009bb.  All of the bolometric LCs have been corrected for foreground and host extinction.  Note that the bolometric LC of SN~1994I and SN~2003dh are in filters $BVR_{c}I_{c}$ and $UBVR_{c}I_{c}$ respectively.}
\label{fig:GRB100316D_Bol_All}
\end{figure}

Figure \ref{fig:GRB100316D_Bol_All} shows the optical \& infrared bolometric light curves of XRF 100316D / SN~2010bh in relation to three spectroscopically-confirmed GRB-SNe: 1998bw, 2003dh \& 2006aj, as well as broad-lined Ic SN~2009bb and Ic SN~1994I, neither of which were accompanied by a GRB-trigger.  All of the bolometric light curves have been determined by integrating the flux in the optical and infrared $UBVR_{c}I_{c}JH$ filters, and have been corrected for foreground and host extinction.  We have constructed the bolometric light curves of SN~1998bw (Galama et al. 1998a; McKenzie \& Schaefer 1999; Sollerman et al. 2002; Patat et al. 2001), SN~2006aj (Campana et al. 2006; Sollerman et al. 2006; Cobb et al. 2006; Ferrero et al. 2006) \& SN~2009bb (Pignatta et al. 2011) from the individual light curves presented in the literature.  We have also incorporated the bolometric data presented in Deng et al. (2005) for SN~2003dh and the bolometric data presented in Richmond et al. (1996) for SN~1994I.  When constructing the bolometric light curve of SN~1998bw, for which very little infrared data are available in the literature, we estimated the amount of flux emitted at IR wavelengths using the behaviour of SNe 2006aj, 2009bb and 2010bh as a proxy.  Note also that the bolometric LCs of SN~1994I and SN~2003dh are in filters $BVR_{c}I_{c}$ and $UBVR_{c}I_{c}$ respectively.  Finally, using the same method that we applied to SN~2010bh in Section \ref{sec:2010bh_2006aj}, we have estimated the amount of flux emitted at UV wavelengths for SN~1998bw by using SN~2006aj as a template/proxy.

As for SN~2006aj, we fit our analytical model to the bolometric light curves of the comparison SNe.  We have used values of the photospheric velocities given in the literature for each event.  The results of our fits are presented in Table \ref{table:Ic_bol_params}, where the quoted errors are statistical only.  

When we compare the values of the nickel mass ($\rm{M_{Ni}}$), ejected mass ($\rm{M_{ej}}$) and explosion energy ($\rm{E_{k}}$) determined from our model with those in the literature, it is seen that our model is able to recover the physical parameters of these events that are determined by more detailed spectral and photometric modeling.  We note that on average, we tend to under-estimate the amount of ejected mass, which in turn under-estimates the explosion energy in these events, apart for SN~2006aj, where we over-estimate the explosion energy.  There are several assumptions that go into our model, such as no asphericity and the central location of the nickel mass during the explosion (that does not mix with the expanding ejecta), as well as a constant value for the optical opacity.  Even if the opacity is approximately constant during the photospheric phase, it may be a slightly different value to that we are using here.  

Finally, the nickel mass determined from our fit of SN~2010bh (Section 5) reiterates the faint nature of SN~2010bh in relation to the other GRB-SNe.  While the nickel mass produced by SN~2010bh is slightly greater than that of SN~1994I, it is less than that seen in other GRB-SNe.   Indeed, for all of the GRB-SNe that have been investigated through spectral and photometric modeling, SN~2010bh appears to be the faintest (i.e. has created the least amount of nickel during the explosion).

In conclusion, the energetics of GRB-SNe and Ic-BL SNe such as SN~2009bb and SN~2002ap are comparable, and are considerably higher than that of other Ibc SNe such as Ic SN~1994I and Ib SN~1999ex.  Additionally, the ejected masses of the GRB-SNe and Ic-BL SNe are commensurate and somwhat higher than some Ibc SNe (e.g. SN~1994I and SN~1999ex), though some Ibc SNe do ejected comparable masses (e.g. SN~2009jf and SN~1999dn).  Finally, the nickel masses of all these events span a range of $\rm 0.07-0.52 M_{\odot}$, with no clear distinction between GRB-SNe, Ic-BL SNe and Ibc SNe.

\subsection{Stretch and Luminosity Factors of the GRB-SNe}

In an attempt to put the luminosity and stretch factors of SN~2010bh (Table \ref{table:2010bh_stretch}) into context, we assembled from the literature all of the photometric data available for most of the previously detected GRB-SNe.  Note that we have not pursued every event but only included those where several measurements were made of the supernova ``bump'' that would allow us to constrain the properties of the associated SN.  

For every GRB-SN event we assume that light is coming from three sources: (1) the afterglow, (2) the supernova and (3) the host galaxy.  First we remove the constant source of flux from the host galaxy, either by using magnitudes determined by image subtraction by the authors (e.g. XRF 020903; Bersier et al. 2006), or by mathematically subtracting the flux from each measurement using observations made of the host galaxy long after the OT had faded below the brightness of the galaxy (e.g. GRB 090618; Cano et al. 2011).

Then, we use the early-time data for each event, and model the behaviour of the afterglow using single or broken power-laws:

\begin{center}
\begin{equation}
\label{equ:afterglow}
 flux\ (t) = f_{o} \times \left(\left(\frac{t}{T_{break}}\right)^{\alpha_{1}} + \left(\frac{t}{T_{break}}\right)^{\alpha_{2}}\right)^{-1}
\end{equation} 
\end{center}

{\noindent}where $f_{o}$ is the flux zero-point, unique to each filter, $t$ is the time since the burst and $T_{break}$ is the time when the power-law changes from temporal index $\alpha_{1}$ to $\alpha_{2}$.  The afterglow parameters $f_{o}$, $T_{break}$, $\alpha_{1}$ and $\alpha_{2}$ were allowed to vary in the fit for each event, and the results of our modeling are listed in Table \ref{table:GRB_afterglow}.  

In all cases the values we derived are similar to those seen in the literature.  For example, for GRB 990712 we find that a single power-law fits both the $V$ and $R$-band data well with a decay constant of $\alpha=-0.95\pm0.02$.  In comparison, Bjornsson et al. (2001) find that the $V$-band data can be fit with either a single or broken power-law, where the value in the former is $\alpha_{V}-0.82\pm0.03$.  They find that the $R$-band photometry can be fit with a single power-law with a slightly steeper decay index of $\alpha_{R} = -0.91\pm0.06$.  Both of these values are fully consistent with our own fit of the optical data. 

Additionally, for GRB 020405 Price et al. (2003) found from their $R$-band photometry a decay constant of $\alpha=-1.41$ when they limited the range of the fit up to 10 days after the burst, while Masetti et al. (2003) find $\alpha=-1.54\pm0.06$ in all optical filters for the same time period.  However, when Price et al. (2003) re-fit the data with a broken power-law, they find a better fit with parameters $\alpha_{1}=-0.94\pm0.25$, $\alpha_{2}=-1.93\pm0.25$, and the time the LC breaks from $\alpha_{1}$ to $\alpha_{2}$ of $T_{break}=1.67\pm0.52$ days ($\chi^{2}/\rm{d.o.f}=35.9/28$).  Similarily, Masetti et al. (2003) find that at times $> 20$ days, $\alpha=-1.85\pm0.15$.  When we restrict our fit to $2-10$ days, we find that a single power-law provides a suitable fit to the data with $\alpha=1.72\pm0.08$, $\chi^{2}/\rm{d.o.f}=6.16/4$.  This value for the temporal index is consistent with the value of $\alpha_{2}$ found by Price et al. (2003) and the value of $\alpha$ found by Masetti et al. (2003) at late times.

\begin{figure*}
\centering
\includegraphics[height=9in,width=6.3in, angle=0]{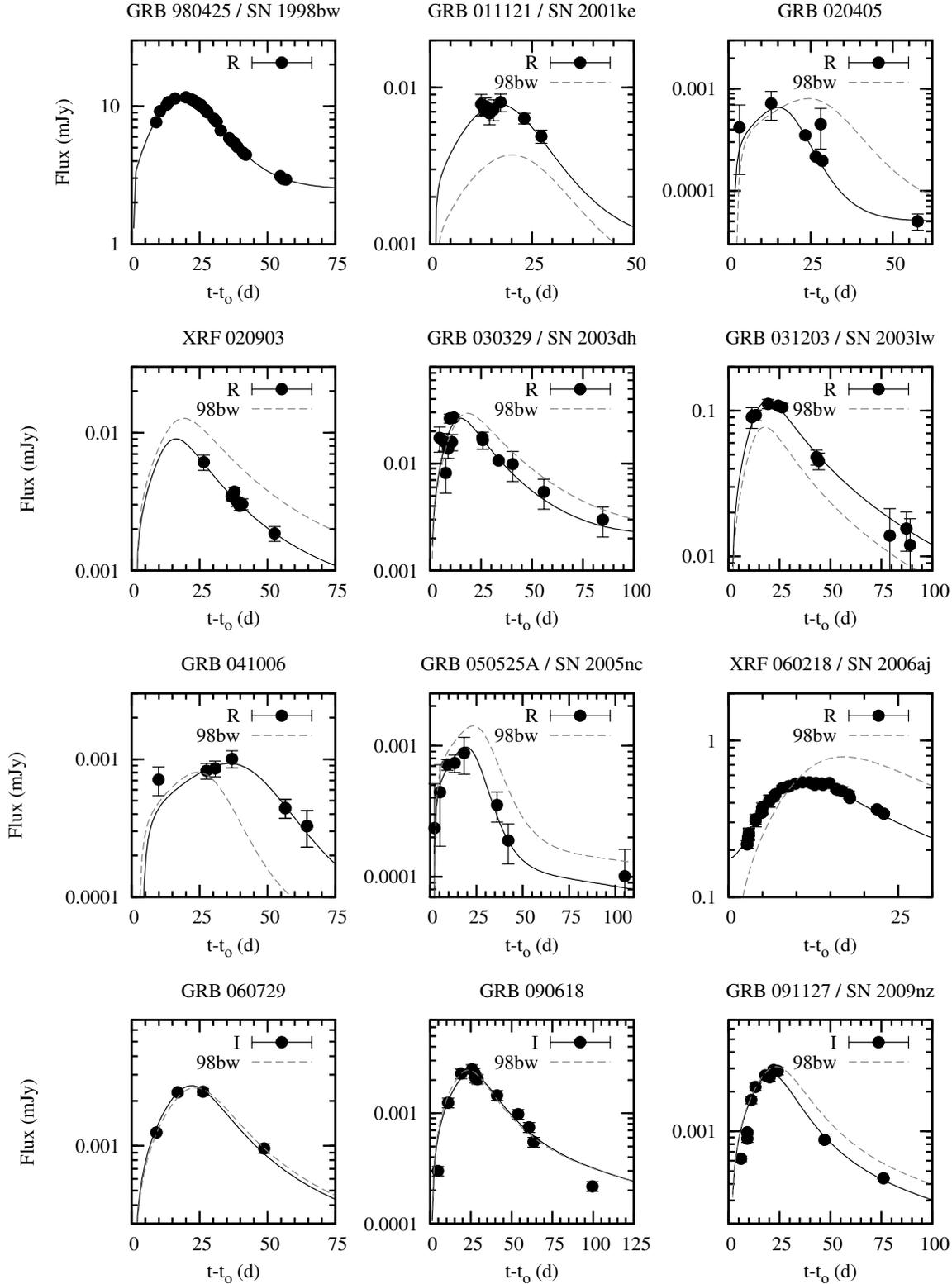}
\caption{Mosaic of GRB-SNe.  For each event the host has been subtracted, either by image subtraction by the original authors or by mathematically subtracting the flux due to the host.  The OT (afterglow and supernova) light curves were then modeled, with the parameters of the afterglow determined and then subtracted out to create host- and afterglow-subtracted ``supernova'' light curves.  Plotted for each event (dashed line) is how SN~1998bw would appear if it occurred at the redshift of the event.  The solid-line is the light curve of SN~1998bw corrected by the luminosity and stretch factors listed in Table \ref{table:GRB_stretch}.  For XRF 060218 / SN~2006aj, we have included an additional power-law component to the SN with index $\alpha=-0.2$, similar to that found by Ferrero et al. (2006). }
\label{fig:GRB_SNe}
\end{figure*}

Once the afterglow parameters were determined, we then subtract Equation (\ref{equ:afterglow}) from the OT (optical transient, e.g. afterglow and supernova) flux light curve to obtain host- and afterglow-subtracted ``supernova'' light curves (e.g. Cano et al. 2011).  The resultant light curves are displayed in Figure \ref{fig:GRB_SNe}.  Note that in Figure \ref{fig:GRB_SNe} for XRF 060218 / SN~2006aj, we have included an additional power-law component of $\alpha=-0.2$, which was fixed in the fit, and is similar to that determined by Ferrero et al. (2006) of $\alpha=-0.20\pm 0.10$.

\begin{figure}
\centering
\includegraphics[scale=0.34, angle=-90]{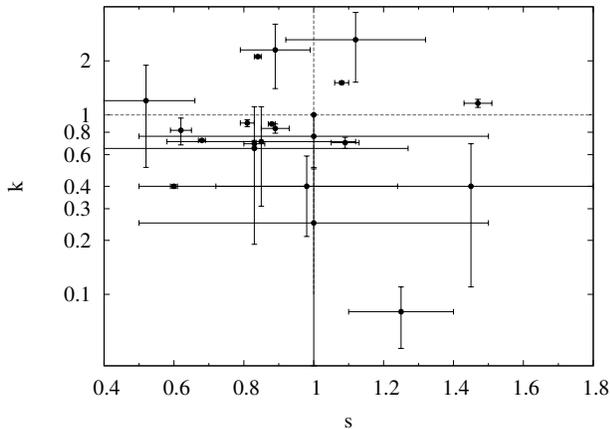}
\caption{Luminosity factor ($k$) versus stretch factor ($s$) in the $R$ band for all of the GRB-SNe in Tables \ref{table:2010bh_stretch} and \ref{table:GRB_stretch}.  We find a correlation/Pearson coefficient for the entire sample of $c=-0.05$, with a $95\%$ confidence interval of ($-0.46,0.38$) for a significance of $p>0.05$.  As the critical value for the Pearson correlation coefficient for $N=22$, $\rm{d.o.f.}=20$ and $p>0.05$ is $0.492$, we can conclude there is not a statistically significant correlation.}
    \label{fig:stretch}
\end{figure}

For each of the host- and afterglow-subtracted ``supernova'' light curves, we determined the stretch and luminosity factor for each GRB-SN in relation to SN~1998bw in various filters, repeating the same process we followed for SN~2010bh (Section 3.5).  The results of our fit are listed in Table \ref{table:GRB_stretch}, with the $R$-band stretch and luminosity factors displayed in Figure \ref{fig:stretch}, and the host- and afterglow-subtracted ``supernova'' light curves displayed in Figure \ref{fig:GRB_SNe}.  

A previous study by Zeh et al. (2004), which was extended by Ferrero et al. (2006) (F06 hereon), also determined the stretch and luminosity factors of GRB-SN in relation to SN~1998bw\footnote{only in the $R$ band} by modeling the SN ``bumps'' seen in the GRB light curves.  For many events where we have determined the stretch and luminosity factors from the afterglow- and host-subtracted ``supernova'' light curves, and F06 from the supernova ``bumps'', we find similar results: e.g. GRB 050525A / SN~2005nc: this paper: $k=0.69\pm0.03$, $s=0.83\pm0.03$; F06\footnote{in the $R$ band and corrected for host-extinction}: $k=0.66^{+0.10}_{-0.08}$, $s=0.77\pm0.04$; and XRF 060218 / SN~2006aj ($V$ band): $k=0.72\pm0.02$, $s=0.65\pm0.02$; F06: $k=0.724\pm 0.007$, $s=0.682\pm0.005$, with similar results for the other filters.

However, for some events we find different properties of the GRB-SN:  for GRB 011121 / SN~2001ke, we find similar stretch factor to F06 (this paper: $s=0.84\pm 0.01$; F06: $s=0.80\pm 0.02$), however we find SN~2001ke to be much brighter once correcting for host-extinction:  (this paper: $k=2.11\pm 0.03$; F06: $k=0.88^{+0.08}_{-0.07}$).  Additionally, for GRB 020405 we find a similar luminosity factor as F06: (this paper: $k=0.82\pm 0.14$; F06: $k=0.90^{+0.15}_{-0.11}$), however we find that the SN evolves more quickly: (this paper: $s=0.62\pm 0.03$; F06: $s=0.97\pm 0.07$).  However, GRB 020405 aside, for every other GRB-SN event we find the same value for the stretch factor as F06 within our respective errorbars. 

We note that when using this method to determine the luminosity factor relative to SN~1998bw, the SN brightness depends on how accurately we have modeled the afterglow.  For instance, if we over-estimate the relative contribution of the afterglow, then we will correspondingly under-estimate the SN brightness.

We find that the stretch factor of each GRB-SN event is approximately the same in each filter (e.g. GRB 990712, GRB 031203, XRF 060218, GRB 060729 \& GRB 090618), though this trend applies less in the bluer filters.  For example, for XRF 060218 / SN~2006aj, we find stretch factors in the $R$ and $I$ filters of $s=0.68\pm0.02$, while in $V$ and $B$ we find slightly smaller values of $s=0.65\pm0.02$ and $s=0.60\pm0.02$ respectively.  

We have investigated whether there might be a correlation between the $k$ and $s$ factors of the GRB-SNe, perhaps something akin to that established for type Ia SNe (i.e. the possibility of using GRB-SNe as ``standard candles'').  We have used all of the $R$-band data, of which there are 18 GRB-SNe.  For GRB/XRFs 021211, 040924, 080319B and 091127, for which no $R$ band data exists, we have assumed that the stretch and luminosity factors in the $V$ band (GRB 021211) and the $I$ band (GRBs 040924, 080319B and 091127) are approximately the same as in the $R$ band.  As already noted for events such as XRF 060218 and GRB 090618, the stretch and luminosity factors in filters $VR_{c}I_{c}$ are approximately the same and differ by only a few percent.  The inclusion of these events brings the total sample size up to $N=22$, and for the entire sample we find a correlation/Pearson coefficient of $c=-0.05$ and a $95\%$ confidence interval of ($-0.46,0.38$) for a significance of $p>0.05$.  As the critical value for the Pearson correlation coefficient for $N=22$, $\rm{d.o.f.}=20$ and $p>0.05$ is $0.492$, we can conclude that a statistically significant correlation does not exist. 

Stanek et al (2005) noted that a relation may exist between light curve shape and luminosity, which the results of F06 could neither confirm nor refute (total sample of $N=13$).  F06 noted that there might be a trend of rising $k$ with rising $s$, however the addition of 9 more GRB-SNe to the total sample shows that it is highly unlikely that such a correlation exists.  When this result is taken in tandem with the results of Drout et al. (2011), who found no statistically-significant correlation between peak luminosity and light-curve decay shapes for a sample of Ibc SNe, indicates that Ibc SNe (including GRB-SNe and Ic-BL SNe) cannot be used as standardizable candles.

\begin{table*}
\begin{center}
\begin{minipage}{105mm}
\caption{Afterglow parameters of the GRB-SNe \label{table:GRB_afterglow}}
\begin{tabular}{rccccc}
\hline
GRB/XRF/SN &  Filter  & $\alpha_{1}$ & $\alpha_{2}$  & $T_{break}$ (days) & $\chi^{2}/\rm{d.o.f}$ \\
\hline
%
011121/2001ke & $R_{c}$ & $-1.70 \pm 0.01$ & - & - & $3.56/4$\\
020405 & $R_{c}$ & $-1.72 \pm 0.16$ & - & - & $6.16/4$ \\
020903 & $R_{c}$ & $-1.87 \pm 0.05$ & - & -  & $1.86/2$\\
041006 & $R_{c}$  & $-0.47 \pm 0.35$ & $-1.47 \pm 0.39$ & $0.30 \pm 0.78$ & $3.24/3$ \\
050525A/2005nc & $R_{c}$  & $-0.89 \pm 0.22$ & $-2.13 \pm 0.30$ & $0.43 \pm 0.35$ & $4.68/3$ \\
060729 & $U$ $^{*}$  & $-0.01 \pm 0.03$ & $-1.65 \pm 0.05$ & $0.75 \pm 0.08$ & $37.99/29$ \\
090618 & $R_{c}$, $i$ $^{\dagger}$  & $-0.65 \pm 0.07$ & $-1.57 \pm 0.07$ & $0.50 \pm 0.11$ & $151.42/113$ \\
\hline
\end{tabular}

\medskip
$^{*}$ Cano et al. (2011) determined the afterglow parameters using the $U$-band data due to the lack of early-time data in $R$ and $I$. \\
$^{\dagger}$ Cano et al. (2011) fit all of the optical data simultaneously. \\
\end{minipage}
\end{center}
\end{table*}

\begin{table*}
\begin{center}
\begin{minipage}{105mm}
\caption{Luminosity ($k$) and Stretch ($s$) Factors of the GRB-SNe with respect to SN~1998bw\label{table:GRB_stretch}}
\begin{tabular}{rccccc}
\hline
GRB/XRF/SN &  Filter & $s$ & $k$ & Ref ($s$ \& $k$) & Ref (Photometry) \\
\hline
970228 & $R$& $1.45\pm0.95$ & $0.40\pm0.29^{*}$ & (1) & - \\
980425 & $BVRi$ & $1.00$ & $1.00$ & - & (4) \\
990712 & $V$ & $0.83\pm0.44$ & $0.65\pm0.46$ & This paper & (5), (6), (7), (8) \\
990712 & $R$ & $0.90\pm0.07$ & $1.45\pm0.20$ & This paper & (5), (6), (7), (8) \\
991208 & $R$ & $1.12\pm0.20$ & $2.62\pm1.10$ & (1) & - \\
000911 & $R$ & $1.40\pm0.32$ & $0.85\pm0.44$ & (1) & - \\
011121/2001ke & $R$ & $0.84\pm0.01$ & $2.11\pm0.03$ & This paper & (9), (10) \\
020405 & $R$ & $0.62\pm0.03$ & $0.82\pm0.14$ & This paper & (11), (12) \\
020903 & $R$ & $0.85\pm0.27$ & $0.71\pm0.40$ & This paper & (13) \\
021211/2002lt & $V$ & $0.98\pm0.26$ & $0.40\pm0.19^{*}$ & (1) & - \\
030329/2003dh & $R$ & $0.81\pm0.02$ & $0.90\pm0.04$ & This paper & (14), (15) \\
031203/2003lw & $R$ & $1.08\pm0.02$ & $1.51\pm0.03$ & This paper & (16), (17), (18) \\
031203/2003lw & $I$ & $1.07\pm0.04$ & $1.29\pm0.04$ & This paper & (16), (17), (18) \\
040924 & $I,z$ & $1.00$ (fixed) & $\approx0.25$ & (2) & - \\
041006 &$ R $& $1.47\pm0.04$ & $1.16\pm0.06$ & This paper & (19) \\
050525A/2005nc & $R$ & $0.83\pm0.03$ & $0.69\pm0.03$ & This paper & (20) \\
050824 & $R$ & $0.52\pm0.14$ & $1.20\pm0.69$ & (23), (24) & - \\
060218/2006aj$^{**}$ & $B$ & $0.60\pm0.01$ & $0.67\pm0.02$ & This paper & (1) \\
060218/2006aj$^{**}$ & $V$ & $0.65\pm0.01$ & $0.72\pm0.01$ & This paper & (1) \\
060218/2006aj$^{**}$ & $R$ & $0.68\pm0.01$ & $0.72\pm0.01$ & This paper & (1) \\
060218/2006aj$^{**}$ & $I$ & $0.68\pm0.01$ & $0.76\pm0.01$ & This paper & (1) \\
060729 & $R$ & $0.89\pm0.04$ & $0.84\pm0.05$ & This paper & (21) \\
060729 & $I$ & $0.94\pm0.02$ & $1.03\pm0.02$ & This paper & (21) \\
080319B & $I$ & $0.89\pm0.10$ & $2.30\pm0.90$ & (3), (25) & - \\
090618 & $R$ & $1.09\pm0.04$ & $0.70\pm0.05$ & This paper & (21) \\
090618 & $I$ & $1.06\pm0.03$ & $0.94\pm0.03$ & This paper & (21) \\
091127 & $I$ & $0.88\pm0.01$ & $0.89\pm0.01$ & This paper & (22) \\
101225A & $R,i,z$ & $1.25\pm0.15$ & $0.08\pm0.03$ & (26) & - \\
\hline
\end{tabular}

\medskip
$^{*}$ Not corrected for host-extinction.\\
$^{**}$ Fit \textbf{excluding} additional decaying power-law component (see text).\\ \ \\
(1) \cite{Ferrero06}, (2) \cite{Soderberg06}, (3) \cite{Tanvir10},  (4) \cite{Galama98}, (5) \cite{Sahu00}, (6) \cite{Christensen04}, (7) \cite{Bjornsson01}, (8) \cite{Hjorth00}, (9) \cite{Garnavich03}, (10) \cite{Kupcu07}, (11) \cite{Bersier03}, (12) \cite{Price03}, (13) \cite{Bersier06}, (14) \cite{Matheson03}, (15) \cite{Deng05}, (16) \cite{Malesani04}, (17) \cite{Mazzali2006}, (18) \cite{Margutti07}, (19) \cite{Stanek05}, (20) \cite{DellaValle06}, (21) \cite{Cano2011}, (22) \cite{Cobb10}, (23) \cite{Sollerman2007}, (24) \cite{Kann2010}, (25) \cite{Bloom2009}, (26) \cite{Thone2011}
\end{minipage}
\end{center}
\end{table*}

\section{Summary}
\label{sec:summary}

We have presented optical and near infrared photometry of XRF 100316D / SN~2010bh obtained on the Faulkes Telescope South, Gemini South and \emph{HST}, with the data spanning from $\rm{t-t_{o}} = 0.5 - 47.3$ days.  It was shown that the optical light curves of SN~2010bh evolve more quickly than the archetype GRB-SN~1998bw, and at a similar rate to SN~2006aj, which was associated with XRF 060218, and non-GRB associated type Ic SN~1994I.  In terms of peak luminosity, SN~2010bh is the faintest SN yet associated (either spectroscopically or photometrically) with a long-duration GRB or XRF, and has a peak, $V$-band, absolute magnitude of $M_{V}=-18.62 \pm 0.08$.  

SN~2010bh appears to be redder than GRB-SNe 1998bw and 2006aj, where the colour curves of SN~2010bh are seen to be redder early on, though at late times the $V-R$ and $R-i$ colour curves matched that of SN~1994I.  The red nature of SN~2010bh is also demonstrated in Figure \ref{fig:GRB100316D_Bol_percentage}, where it is seen that more of the bolometric flux is emitted at infrared wavelengths (and less at optical wavelengths) than in the broad-lined Ic SN~2009bb. 

We also gave evidence of the detection of light coming from the shock break-out at $\rm{t-t_{o}}=0.598$ days.  The brightness of the $B$-band light curve at this epoch, as well as the shape of the optical SED, which with a spectral power-law index of $\beta=+0.94\pm0.05$ (which is harder than is expected for synchrotron radiation), implies that the source of light at this epoch is not synchrotron in origin and is likely coming from the shock-heated, expanding stellar envelope.  

We then applied a simple physical model to the bolometric light curve of SN~2010bh.  When we include all photometry in the optical and infrared regime (in the $\rm 3,000\AA-16,600 \AA$ wavelength range) we find physical parameters of $\rm{M_{Ni}}=0.10 \pm 0.01 M_{\odot}$, $\rm{M_{ej}}=2.24 \pm 0.08 M_{\odot}$, $\rm{E_{k}} = 1.39 \pm 0.06  \times 10^{52}$ erg.  The faint nature of SN~2010bh becomes again apparent when the nickel mass synthesized during the explosion is compared with other GRB-SNe such as SN~1998bw, where it is believed that $\rm{M_{Ni}}\approx 0.4 M_{\odot}$ was created in the explosion.  Indeed SN~2010bh synthesized only a marginally larger amount of nickel than local type Ic SN~1994I ($\rm{M_{Ni}}\approx 0.07 M_{\odot}$).  The investigation of SN~2010bh in relation to the general population of type Ic SNe has once again shown that type Ic SNe are a very heterogeneous class of supernovae, spanning a wide range of luminosities and ejected masses.

Finally, we assembled from the literature all of the available photometry of previously detected GRB-SNe.  For each of these events we assumed that light is coming from three sources: from the host galaxy, the afterglow and the supernova.  First we removed the constant flux due to the host galaxy, then we modeled the optical afterglows, and then subtracted the light due to the afterglow, which resulted in host- and afterglow-subtracted ``supernova'' light curves (Figure \ref{fig:GRB_SNe}).  We then created synthetic light curves of SN~1998bw as it would appear if it occurred at the redshift of each given GRB-SNe.  We then compared the brightness and shape of the GRB-SNe with that of the shifted LCs of SN~1998bw, and derived stretch ($s$) and luminosity ($k$) factors for each GRB-SNe.  The results of our method, when compared to previous studies performed by Ferrero et al. (2006) showed similar values for the stretch and luminosity factors, as well as including more events than the previous studies.  When we checked for the possibility of a correlation between $k$ and $s$ in the $R$-band data, we find an insignificant correlation/Pearson coefficient of $c=-0.05$, suggesting that it is highly unlikely that such a relation exists.

\acknowledgments

This work was supported partially by a Science and Technology Facilities Council (STFC) (UK) research studentship (ZC).  Support for program \# 11709 was provided by NASA through a grant from the Space Telescope Science Institute, which is operated by the Association of Universities for Research in Astronomy, Inc., under NASA contract NAS 5-26555.  AG acknowledges funding from the Slovenian Research Agency and from the Centre of Excellence for Space Science and Technologies SPACE-SI, an operation partly financed by the European Union, European Regional Development Fund and Republic of Slovenia, Ministry of Higher Education, Science and Technology.  JG is supported by the Spanish programs AYA2007-63677, AYA2008- 03467/ESP, and AYA2009-14000-C03-01.

{\it Facility:} \facility{HST (WFC3)},
\facility{FTS},
\facility{Gemini:South (GMOS)}

{}

\end{document}